\documentclass[conference, onecolumn]{IEEEtran}

\usepackage{graphicx}
\usepackage{hyperref}
\usepackage[ruled,vlined]{algorithm2e}
\usepackage{amsmath}
\usepackage{amsfonts}
\usepackage{subcaption}
\usepackage{booktabs}
\begin{document}
\title{A Novel Data Encryption Method Inspired by Adversarial Attacks}

\author{\IEEEauthorblockN{Praveen Fernando}
\IEEEauthorblockA{Department of Computer and Information Technology\\
Purdue University\\
West Lafayette, IN\\
Email: ferna159@purdue.edu}
\and
\IEEEauthorblockN{Jin Wei-Kocsis}
\IEEEauthorblockA{Department of Computer and Information Technology\\
Purdue University\\
West Lafayette, IN\\
Email: kocsis0@purdue.edu}}

\maketitle

\begin{abstract}
Due to the advances of sensing and storage technologies, a tremendous amount of data becomes available and, it supports the phenomenal growth of artificial intelligence (AI) techniques especially, deep learning (DL), in various application domains. While the data sources become valuable assets for enabling the success of autonomous decision-making, they also lead to critical vulnerabilities in privacy and security. For example, data leakage can be exploited via querying and eavesdropping in the exploratory phase for black-box attacks against DL-based autonomous decision-making systems. To address this issue, in this work, we propose a novel data encryption method, called AdvEncryption, by exploiting the principle of adversarial attacks. Different from existing encryption technologies, the AdvEncryption method is not developed to prevent attackers from exploiting the dataset. Instead, our proposed method aims to trap the attackers in a misleading feature distillation of the data.  To achieve this goal, our AdvEncryption method consists of two essential components: 1) an adversarial attack-inspired encryption mechanism to encrypt the data with stealthy adversarial perturbation, and 2) a decryption mechanism that minimizes the impact of the perturbations on the effectiveness of autonomous decision making. In the performance evaluation section, we evaluate the performance of our proposed AdvEncryption method through case studies considering different scenarios.
\end{abstract}

\section{Introduction}
Due to the advances in sensing and computing, DL technologies are widely used in various application fields, including self-driving vehicles \cite{Grigorescu2020ADriving}, automated language translation \cite{Devlin2019BERT:Understanding}, and robotics operations \cite{Levine2018LearningCollection}. With the rise of such deep learning technologies, data sources have become valuable assets for enabling the success of autonomous decision-making, which have a high vulnerability to potential attacks. For example, successful feature distillation on critical data will give an attacker the capability to understand the internal workings of the associated DL-based autonomous decision making models. Based on the knowledge distilled from the data, the attacker can launch attacks, such as evasion attacks, with a higher success rate \cite{Carlini2017TowardsNetworks, Carlini2018AudioSpeech-to-Text}.

Various encryption mechanisms have been developed to enhance data security and privacy. As one type of the most widely used techniques, key-based encryption methods can be generally classified into two groups: symmetric key encryption methods and asymmetric key encryption methods. In symmetric key encryption, an identical pair of keys are used for encryption and decryption \cite{Kumar2006FundamentalsCryptography}. Advanced Encryption Standard (AES) \cite{Daemen2001SpecificationAES} is an example of symmetric key encryption method. In asymmetric key encryption, a public key is used to encrypt data, whereas a private key is used to decrypt it \cite{Koblitz2004ACryptosystems}. Rivest–Shamir–Adleman (RSA) \cite{Rivest1978ACryptosystems} is an example of an asymmetric key encryption method. Homomorphic Encryption (HE) is another form of encryption that enables mathematical operations to be performed directly on the encrypted data. HE has been successfully applied in various classification and regression tasks \cite{Vizitiu2020ApplyingData, Gilad-Bachrach2016CryptoNets:Accuracy, Hesamifard2017CryptoDL:Data}. 

One common property of the aforementioned encryption techniques is that the encrypted data do not carry any meaningful domain-specific semantics of the underlying data distribution and thus are easily distinguishable from the plaintext data. Therefore, the existence of these encryption schemes can be easily recognizable by an attacker who can decide to allocate their effort to identify other attack vectors to reduce the cost. In other words, these encryption techniques prevent attackers from exploiting the vulnerabilities of a dataset, which can result in the change of attack vectors. It would be good if we can trap the attackers in misleading data exploitation so that the attackers cannot launch other attack vectors. Inspired by this essential idea, we develop an innovative encryption method, AdvEncryption, which contains an adversarial attack-inspired encryption mechanism that is used to encrypt data with stealthy adversarial perturbations. By doing so, the attackers can be tricked to consider encrypted data as plaintext data and exhaust their efforts to implement unsuccessful feature distillation on the encrypted data. Furthermore, to minimize the impact of the perturbation on the effectiveness of autonomous decision making, we also design a decryption mechanism in our proposed AdvEncryption method. 

The rest of the paper is organized as follows. In Section II, we will discuss the problem formulation. In Section III, we will introduce the proposed AdvEncryption method. In sections IV and V, the performance evaluation and conclusions will be presented, respectively. 

\section{Problem Formulation}

\begin{figure}[!t]
	\centering
    \includegraphics[width=0.7\textwidth , trim=0 20 0 40,clip]{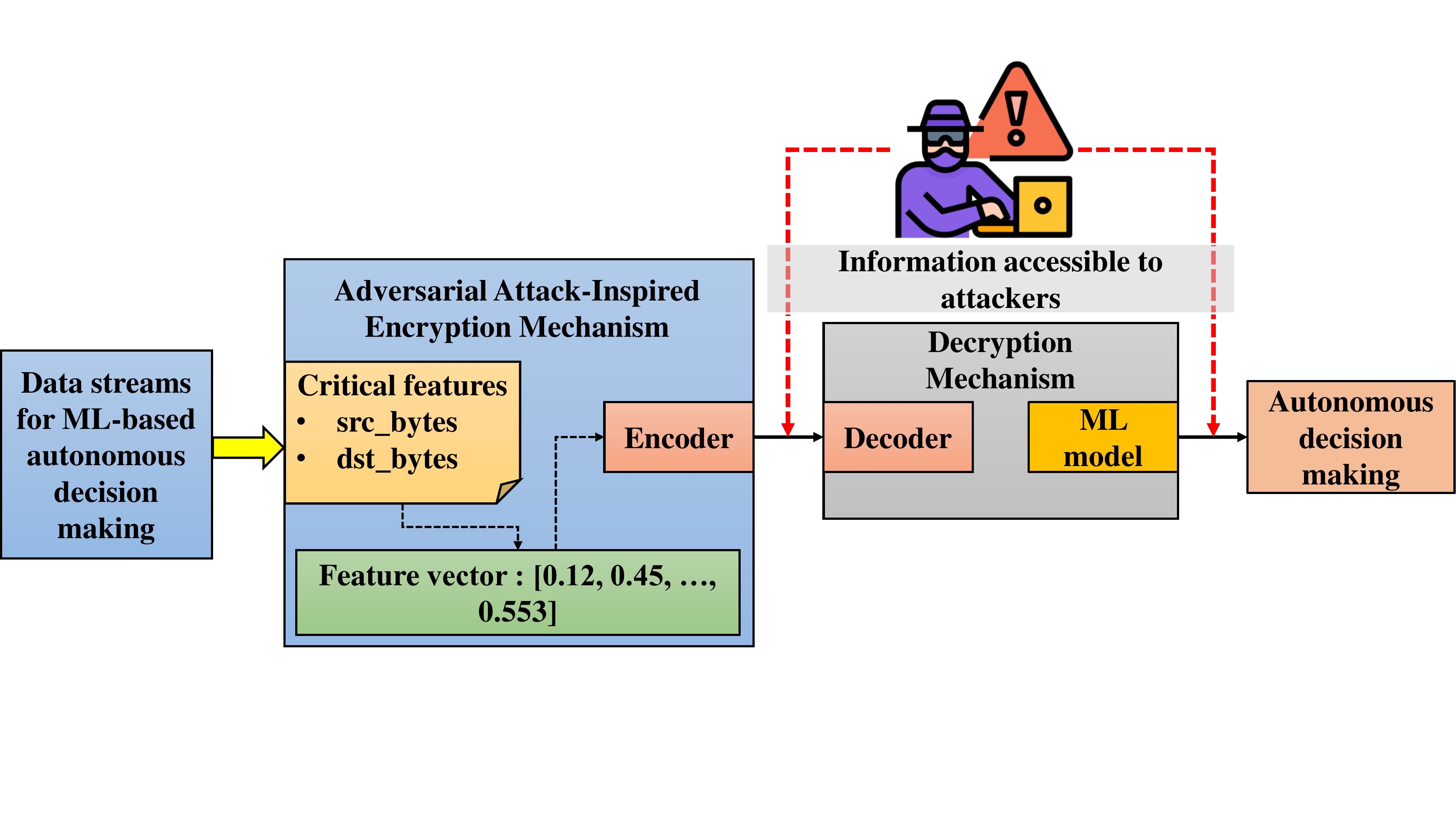}
    \caption{Operation overview of our proposed AdvEncryption method.}
    \label{fig:system_overview}
\end{figure}

The overview of our proposed data encryption method, AdvEncryption, is illustrated in Fig. \ref{fig:system_overview}. As illustrated, we assume that the attackers aim to launch eavesdropping and querying attacks on autonomous decision-making procedures to obtain critical information of the associated data streams. Our proposed method does not limit the maximum number of queries carried out by the attackers. For simplicity, in this paper, we only consider autonomous decision making systems that can be formulated as classification procedures. Our proposed AdvEncryption method mainly consists of two components: 1) adversarial attack-inspired encryption mechanism, and 2) decryption mechanism. The proposed encryption mechanism consists of two main steps. First, the critical features of the data streams for ML-based autonomous decision making are identified. For example, as illustrated in Fig. \ref{fig:system_overview}, in a cybersecurity scenario, the critical features can include source bytes, destination bytes, and protocol type. Second, the identified critical features are encrypted with stealthy adversarial perturbations to mislead the attackers during critical feature distillation. The essential idea is to develop an encoder to realize the mapping between the plaintext data $x^i$ and the encrypted data $x^m$: $x^i \rightarrow x^m$, where $i$ is the ground-truth label of the plaintext data $x^i$ and $m$ is a randomly-selected unified fake label for the encrypted data $x^m$. The proposed decryption mechanism is designed to minimize the impact of the introduced perturbations on the effectiveness of autonomous decision making. To achieve this goal, a decoder is developed to decrypt data $x^m$ and obtain $\tilde{x}^i$ that is used for autonomous decision making. 

\begin{figure}[b]
	\centering
    \includegraphics[width=0.7\textwidth , trim={4 20 4 50},clip]{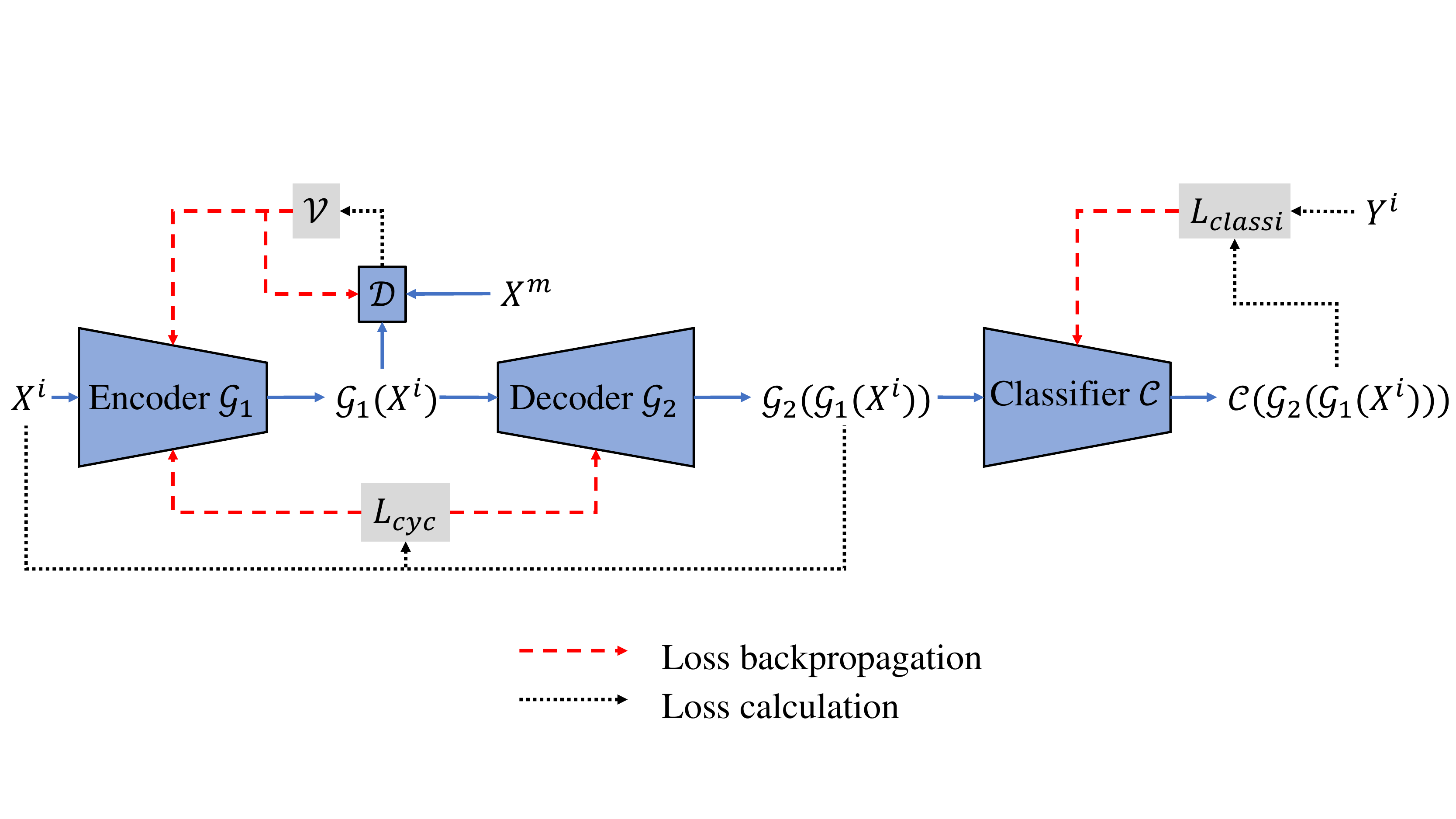}
    \caption{Component-level realization of our proposed AdvEncryption method.}
    \label{fig:system_components}
\end{figure}

\section{Proposed AdvEncryption Method}

To elaborate on our proposed method, we consider that the autonomous decision-making procedure can be formulated as a $K$-class classification procedure. Let $X^i=\{x^i\}$ denote the collection of plaintext data $x^i$ belonging to the class with a ground-truth label $i$, where $i \in \{1, 2, \dots, K\}$. Let $m$ be the randomly selected unified fake label, which is also the targeted label of the encrypted data in the encrypted data collection $X^m = \{x^m\}$. 

By converting all the plaintext data with their ground-truth labels into encrypted data with a unified fake label, we are able to maximize the difficulties of an attacker in feature distillation via identifying a mapping between the encrypted data and the decision-making output and realize the misleading purpose. The component-level realization of our proposed method is illustrated in Fig. \ref{fig:system_components}. As shown in the figure, the design of our proposed AdvEncryption mainly consists of an encoder $\mathcal{G}_1$, a decoder $\mathcal{G}_2$, a discriminator $\mathcal{D}$, and a classifier $\mathcal{C}$. 

To mislead the attacker to consider the encrypted data as plaintext data, the generated adversarial perturbation for encryption needs to be stealthy so that the encrypted data look like plaintext data. To achieve this goal, we exploit GAN technique \cite{Goodfellow2014GenerativeNetworks} to design an encoder $\mathcal{G}_1: X^i \rightarrow X^m$. The encoder is optimized via interacting with a discriminator $\mathcal{D}$ in a GAN training, which can be formulated as follows:

 \begin{equation}
 \begin{aligned}
 \min_{\mathcal{G}_1}\max_{\mathcal{D}} V(\mathcal{G}_1, \mathcal{D}, X^m, X^i) = \mathbb{E}_{x^m \sim p_{data}(x^m)}[log(\mathcal{D}(x^m))]
 + \mathbb{E}_{x^i \sim p_{data}(x^i)}[log(1 - \mathcal{D}(\mathcal{G}_1(x^i)))]
 \end{aligned}
 \label{eq:gan_loss}
\end{equation}

As described in Eq. (\ref{eq:gan_loss}), $\mathcal{D}$ is designed to discriminate between the features of encrypted data $x^m \in X^m$ and the generated features by the encoder $\mathcal{G}_1(X^i)$. The encoder aims to optimize the generation of encrypted data $x^m$ with realistic features that can bypass the discriminator.

While encrypting the data streams to defend against the attackers, we also need to ensure the autonomous decision-making procedure is not interrupted by the data encryption. One possible solution is to directly use the encoded features $\mathcal{G}_1(X^i)$ for the decision making that is formulated as a classification procedure in our paper. However, this solution is not practical due to the possible mode collapse at the GAN-based encoding stage. As stated by Goodfellow \cite{Goodfellow2016NIPSNetworks}, mode collapse occurs when the encoder maps multiple features into the same output point. This adversely affects the subsequent classification task, where the same encoded feature will map to multiple classification labels. To prevent mode collapse, we develop a decryption mechanism by designing a decoder $\mathcal{G}_2$. To realize this goal, we integrate cycle consistency to ensure that the encoded features are able to appropriately decoded back into their original features by exploiting the cycle-GAN technique \cite{Zhu2017UnpairedNetworks}. The cycle-consistency loss of the encoder and decoder can be formulated as follows:

 \begin{equation}
 \begin{aligned}
 \mathcal{L}_{cyc}(\mathcal{G}_1, \mathcal{G}_2, X^i) = \mathbb{E}_{x^i \sim p_{data}(x^i)}||\mathcal{G}_2(\mathcal{G}_1(x^i)) - x^i||_1
 \end{aligned}
\end{equation}

The integrated cycle consistency effectively eliminates one-to-many mappings between the features and labels used in the classification procedure. The output of the decoder $\mathcal{G}_2$, $\tilde{X}^i = \mathcal{G}_2(\mathcal{G}_1(X^i))$, is the decrypted data collection used for the autonomous decision making (i.e. classification procedure in our work). In other words, the decision making on classification is conducted based on $\tilde{X}^i$.  Let $\mathcal{E} (\cdot)$ denote the cross-entropy criterion. We can formulate the loss function for optimizing the classification procedure as follows:

\begin{equation}
 \begin{aligned}
 \mathcal{L}_{classi}(\mathcal{C}, X^i, Y^i) = \mathcal{E} (\mathcal{C}(\mathcal{G}_2(\mathcal{G}_1(X^i))), Y^i)
 \end{aligned}
\end{equation}

Based on the analysis stated above, we can formulate the overall loss function to optimize our AdvEncryption method shown in Fig. \ref{fig:system_components} as follows:

\begin{equation}
 \begin{aligned}
\mathcal{L}_{tot}(\mathcal{G}_1, \mathcal{G}_2, \mathcal{D}, \mathcal{C}, X^m, X^i, Y^i) = V(\mathcal{G}_1, \mathcal{D}, X^m, X^i) 
+ \lambda \cdot \mathcal{L}_{cyc}(\mathcal{G}_1, \mathcal{G}_2, X^i)
+ \mathcal{L}_{classi}(\mathcal{C}, X^i, Y^i) 
 \end{aligned}
\end{equation}

\noindent
where $\lambda$ is a hyperparameter to control the importance of cyclic loss in the overall loss. Based on the overall loss, we formulate the final objective function as follows, which results in optimal $\mathcal{G}_1^*$, $\mathcal{G}_2^*$, and $\mathcal{D}^*$ and  $\mathcal{C}^*$.

\begin{equation}
 \begin{aligned}
\mathcal{G}_1^*, \mathcal{G}_2^*, \mathcal{D}^*,  \mathcal{C}^* = arg \min_{\mathcal{G}_1, \mathcal{G}_2, \mathcal{C}} arg \max_{\mathcal{D}} \mathcal{L}_{tot} (\mathcal{G}_1, \mathcal{G}_2, \mathcal{D}, \mathcal{C}, X^m, X^i, Y^i)
 \end{aligned}
\end{equation}

\section{Performance Evaluations}

\subsection{Evaluation Strategy}

Considering the essential principles of our proposed AdvEncryption method, we evaluate its effectiveness by measuring the success rate of the attackers to reconstruct a substitute model by leveraging feature distillation on the data streams used in the autonomous decision making procedure. The hypothesis of our evaluation strategy design is: the better performance the proposed AdvEncryption has, the less information can be distilled from the data, and thus the lower effectiveness of the substitute model can be obtained for emulating autonomous decision making procedure.

In our evaluation strategy, the attacker is assumed to construct the substitute model via two types of methods: 1) gradient-based method \cite{Li2018Query-EfficientLearning} that is detailed in Algorithm \ref{tab:subst_train_algo_grad}, and 2) GAN-based method \cite{Zhou2020DaST:Attacks} that is detailed in Algorithm \ref{tab:subst_train_algo_gan}. For gradient-based method, the $S_{add}$ in Algorithm \ref{tab:subst_train_algo_grad} is obtained by leveraging Fast Gradient Sign Method (FGSM) \cite{Goodfellow2015ExplainingExamples}, Fast Gradient Value method (FGV) \cite{Rozsa2016AdversarialGeneration}, and Jacobian-based dataset augmentation \cite{Papernot2017PracticalLearning}, respectively, with the following calculations.

\begin{equation}
\begin{aligned}
\text{FGSM : }S_{add} &= \{x + \lambda \cdot sign(\nabla_x J(\theta, x, \widetilde{O}(x))) | x \in S_{\rho}\} \\
\text{FGV : }S_{add} &= \{x + \lambda \cdot \nabla_x J(\theta, x, \widetilde{O}(x)) | x \in S_{\rho}\} \\
\text{Jacobian-based : }S_{add} &= \{x + \lambda \cdot sign(J_D[\widetilde{O}(x)]) | x \in S_{\rho}\} \\
\end{aligned}
\label{eq:grad_methods}
\end{equation}

\begin{algorithm}[!t]
\SetAlgoLined
\textbf{INPUT} : target oracle $\widetilde{O}$, a maximum number $\rho_{max}$ of training epochs, and an initial training set $S_0$.

\textbf{OUTPUT} : a trained substitute model $\mathcal{F}$.

Define architecture $\mathcal{F}$;

\For{$\rho = 0; \rho < \rho_{max}; \rho ++$}                  
{
    $D \leftarrow \{ (x, \widetilde{O}(x)) | x \in S_{\rho}\}$;\\
	train $\mathcal{F}$ with $D$;\\
	craft $S_{add}$ using $S_{\rho}$;\\
	$S_{\rho + 1} \leftarrow S_{\rho} \bigcup S_{add}$
}

\caption{Substitute neural network training (Gradient-based method)}
\label{tab:subst_train_algo_grad}
\end{algorithm}

\begin{algorithm}[!t]
\SetAlgoLined
\textbf{INPUT} : target oracle $\widetilde{O}$, a maximum number $\rho_{max}$ of training epochs, and an initial training set $S_0$.

\textbf{OUTPUT} : a trained substitute model $\mathcal{F}$.

Define architecture $\mathcal{F}$ and $\Psi_1, \dots, \Psi_K$, where $K$ is the number of classes;

\For{$\rho = 0; \rho < \rho_{max}; \rho ++$}                  
{
	train $\mathcal{F}$ with $\{ (x, \widetilde{O}(x)) | x \in S_0\}$;\\
	Generate $m$ examples each from $\Psi_1, \dots, \Psi_K$. For $\Psi_j$ belonging to class $j$, generated examples can be denoted as $\{x_i^j\}_{i=1}^m$;\\
	Let $\overline{X} = \bigcup \limits_{j=1}^{K} \{x_i^j\}_{i=1}^m$;\\
	\textbf{Update the substitute model :}\\
	$L_\mathcal{F} = \mathcal{E}(\mathcal{F}(\overline{X}), \widetilde{O}(\overline{X}))$;\\
	\textbf{Update the generator models :}\\
	\For{$j = 1; j \leq K; j ++$}
	{
	    $L_{\Psi_j}= \mathcal{E}(\mathcal{F}(\{x_i^j\}_{i=1}^m), j)$;
	}
}

\caption{Substitute neural network training (GAN-based method)}
\label{tab:subst_train_algo_gan}
\end{algorithm}

\subsection{Training and attack generation}

We consider three different scenarios in our evaluation strategy: (1) cyber attack detection, (2) handwritten digit recognition, and (3) facial recognition. In each scenario, we model the ML-based autonomous decision making via a neural network (NN)-based classifier, which is used as the baseline for performance evaluation. The NN-based classifier is optimized via offline training with the raw training data and corresponding ground-truth labels. Furthermore, we define two metrics to determine the stopping condition of the encoder and decoder training in our AdvEncryption method: 1) encryption efficiency that is the percentage of encoded testing data identified by the baseline NN-based classifier as the targeted fake label, and 2) decryption efficiency that is the element-wise similarity between the plaintext data and the corresponding decoded data.

Additionally, the success rate of the attackers is formulated based on the effectiveness of the substitute model constructed by the attackers based on the feature distillation on the data streams. Specifically, the success rate can be calculated based on the alignment between the outputs of the reconstructed substitute model and the oracle of the defender associated with the same input data. 

All the experiments in the paper are carried out using the setup mentioned in Table \ref{tab:comp_infa}. 

\begin{table}[!t]
	\caption{Computing infrastructure used in the experiments}
   \label{tab:comp_infa}
  	\centering
  	\begin{tabular}{p{0.3\linewidth} p{0.5\linewidth}}
    \toprule
    Resource & Resource description\\
    \midrule
    Operating system & Ubuntu 20.04\\
    \midrule
    Python version & 3.8.10\\
    \midrule
    GPU type & Nvidia Tesla V100\\
    \midrule
    Cuda version & 11.4\\
    \midrule
    Major Python libraries and versions& numpy (1.19.5), matplotlib (3.3.4), scikit-learn (0.24.1), torch (1.8.1), torchvision (0.9.1)\\
    \bottomrule
  	\end{tabular}
\end{table}

\subsection{Scenario 1: Cyber Attack Detection}

\subsubsection{Dataset preparation}

In this scenario, we use the CICDDoS dataset \cite{Sharafaldin2019DevelopingTaxonomy} that consists of network flows that can be benign or belong to one of the 12 Distributed Denial of Service (DDoS) attacks. By randomly sampling the dataset, we can obtain a binary-class dataset, in which the class of benign data flows has label 0 and the class of DDoS-attack data flows has label 1. Additional details of the binary-class dataset is shown in Table \ref{tab:cicddos_2019}. We randomly choose label 1 as the targeted fake label.

\begin{table}[!t]
	\caption{Numbers of training and testing data samples of binary-class dataset based on CICDDoS-2019 dataset}
   \label{tab:cicddos_2019}
  	\centering
  	\begin{tabular}{p{0.2\linewidth} p{0.2\linewidth} p{0.2\linewidth} p{0.2\linewidth}}
    \toprule
    Dataset & Total samples & Benign samples & Malicious samples\\
    \midrule
    Training & 26534 & 13267 & 13267\\
    Testing & 8844 & 4422  & 4422\\
    \bottomrule
  	\end{tabular}
\end{table}

\subsubsection{Training details}

In this case study, we use fully connected NNs to realize the AdvEncryption method and the baseline classifier. We use an Adam optimizer \cite{Kingma2015Adam:Optimization}, to train these neural networks. We use a learning rate of $10^{-5}$ for all the NNs. Additionally, we chain together the encoder and the decoder in a single optimizer to optimize both of them simultaneously. Extracted plaintext features in the study have a dimension of 81. In NNs used for classifiers, there are five sequential linear layers. The first four linear layers have an output size of 512 and are followed by ReLU activation. The last layer has an output size of 2. There is a dropout layer positioned after the second ReLU activation. The encoder and decoder of the AdvEncryption method are identical and have three linear layers. The first two layers are followed by a LeakyReLU activation function and the last layer is followed by a sigmoid activation function. Furthermore, the first two layers have an output size of 512, and the last layer has an output size of 81. Both  LeakyReLU activations are followed by a dropout layer. The discriminator NN of the AdvEncryption method has a structure similar to the encoder, and the decoder. The only difference is that, its final layer has an output size of 1. 

The training and testing accuracies of the NN-based classifier with and without using our proposed AdvEncryption, respectively, are shown in Table \ref{tab:cicddos_2019_acc}. From Table \ref{tab:cicddos_2019_acc}, it can be observed that, by integrating AdvEncryption, the accuracy of the NN-based classifier is slightly reduced. This is caused by the additional noise introduced in encryption and decryption mechanisms in our AdvEncryption. However, the training and testing accuracies remain at an efficiently high level. Additionally, our proposed method can achieve encryption and decryption efficiencies of over $98\%$.

\begin{table}[!b]
	\caption{Training and testing accuracies (CICDDoS-2019)}
  \label{tab:cicddos_2019_acc}
  	\centering
  	\begin{tabular}{p{0.2\linewidth} p{0.3\linewidth} p{0.3\linewidth}}
    \toprule
    & Training accuracy & Testing accuracy \\
    \midrule
    Without AdvEncryption & 99.94347\% & 99.81909\%\\
    With AdvEncryption    & 99.79649\% & 99.60425\%\\
    \midrule
    \multicolumn{3}{l}{Encryption efficiency : 100.00000\%}\\
    \midrule
    \multicolumn{3}{l}{Decryption efficiency : 98.04072\%}\\
    \bottomrule
  	\end{tabular}
\end{table}

\subsubsection{Performance evaluation}

\begin{figure}[!t]
    \centering
    \begin{subfigure}[t]{0.3\textwidth}
        \centering
        \includegraphics[width=\textwidth]{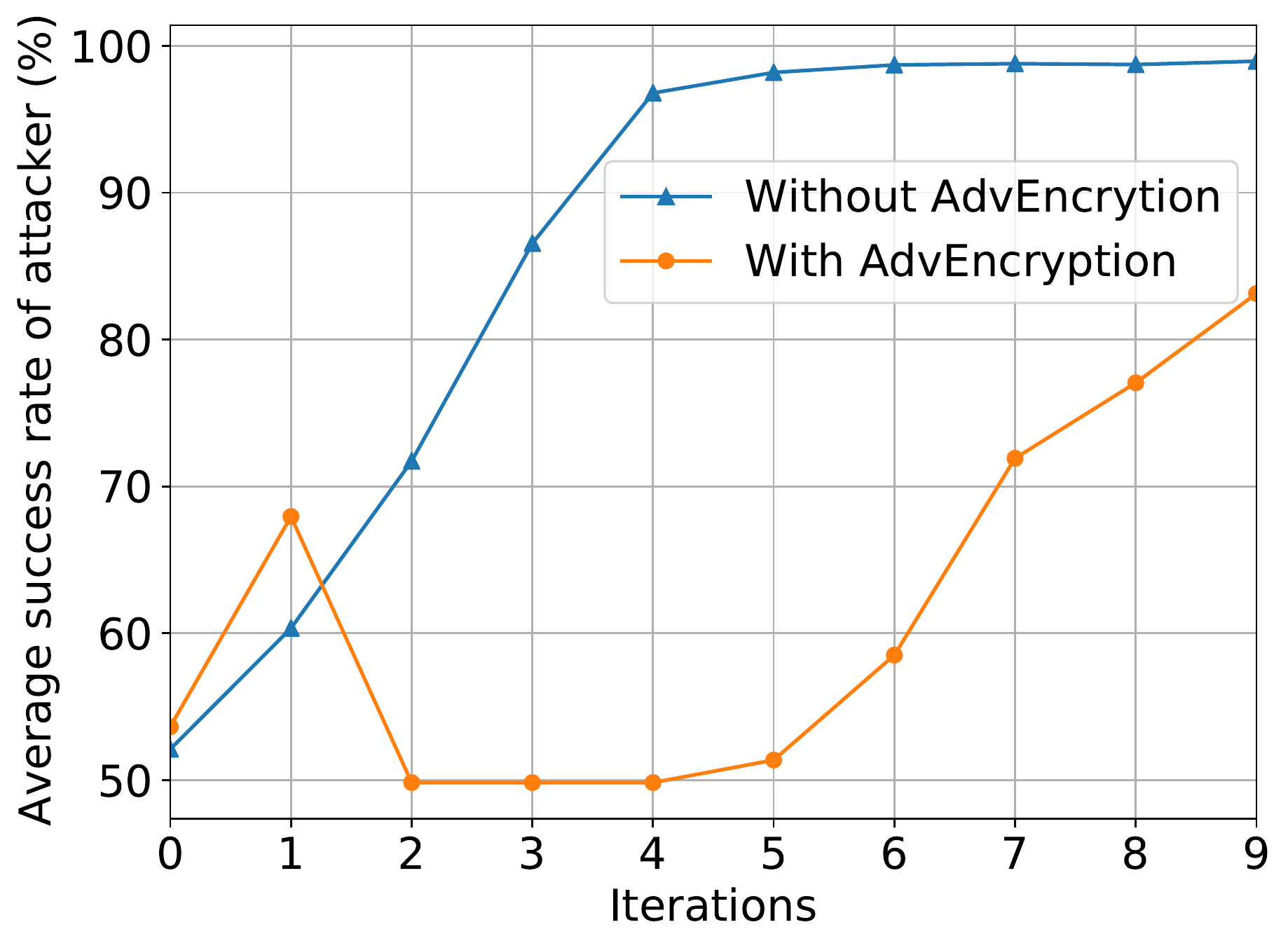}
        \caption{Jacobian-based method}
        \label{fig:jac_cicddos}
    \end{subfigure}
    \hfill
    \begin{subfigure}[t]{0.3\textwidth}
        \centering
        \includegraphics[width=\textwidth]{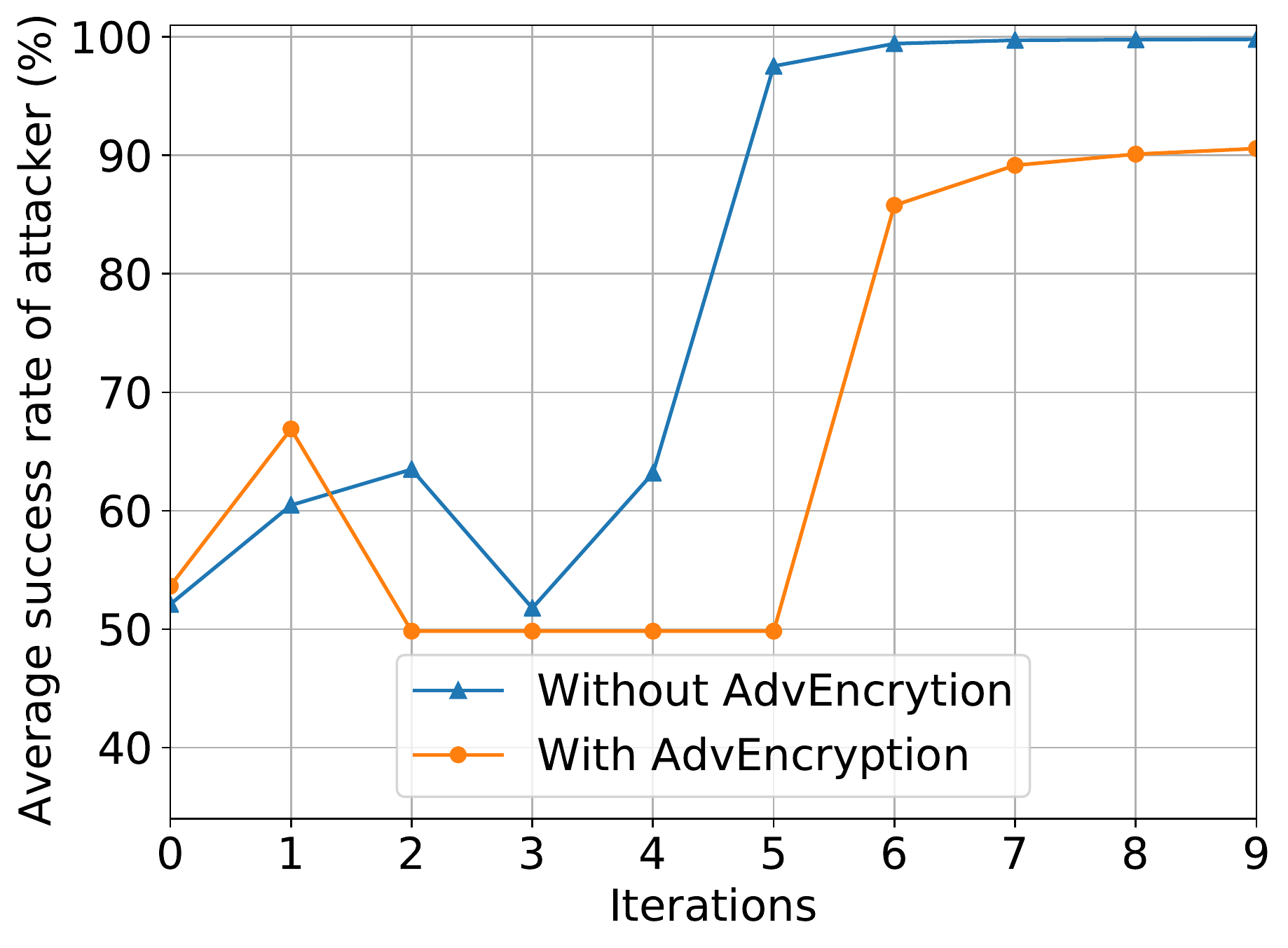}
        \caption{FGSM-based method}
        \label{fig:fgsm_cicddos}
    \end{subfigure}
    \hfill
    \begin{subfigure}[t]{0.3\textwidth}
        \centering
        \includegraphics[width=\textwidth]{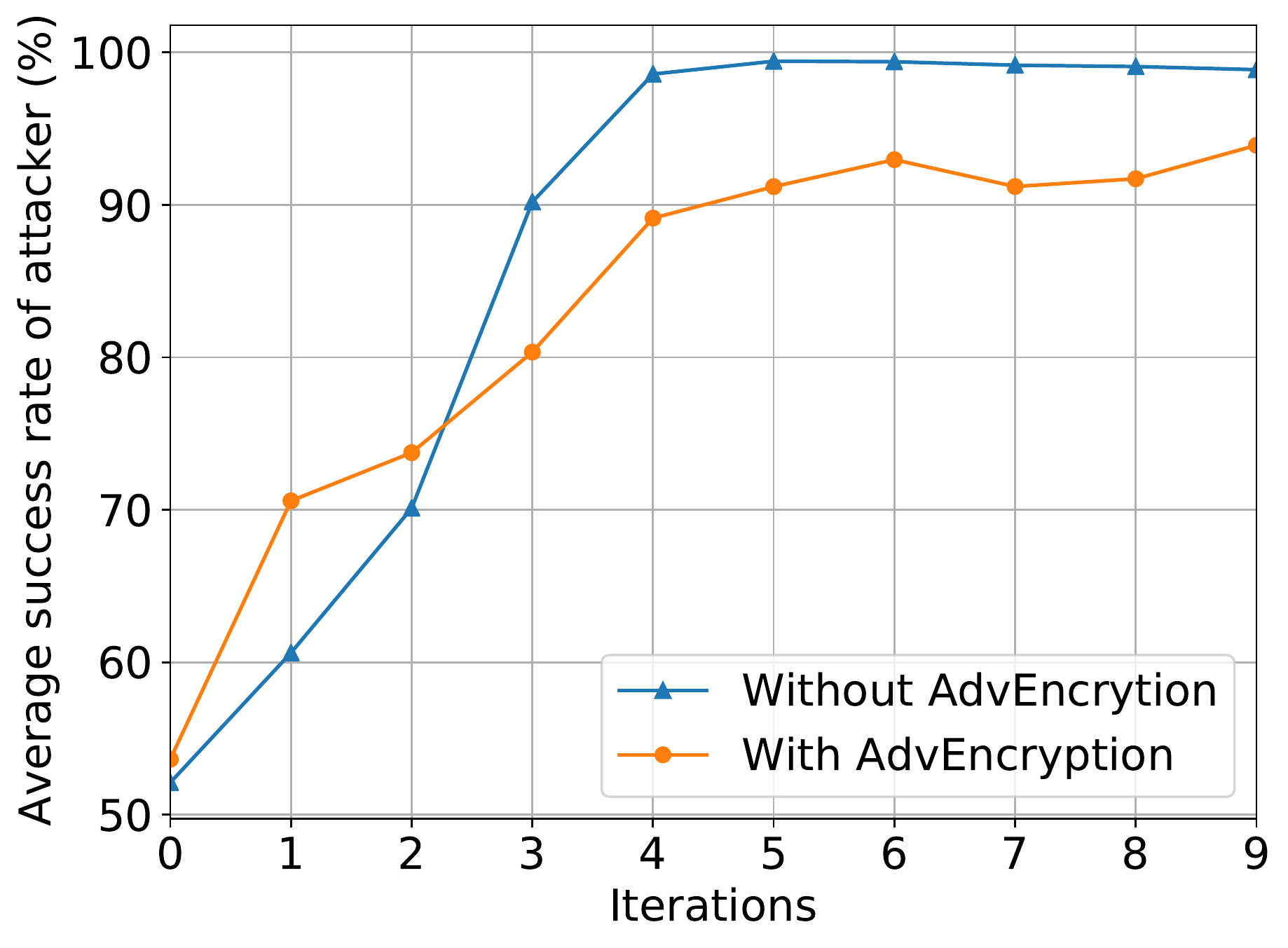}
        \caption{FGV-based method}
        \label{fig:fgv_cicddos}
    \end{subfigure}
    \hfill
    \begin{subfigure}[t]{0.3\textwidth}
        \centering
        \includegraphics[width=\textwidth]{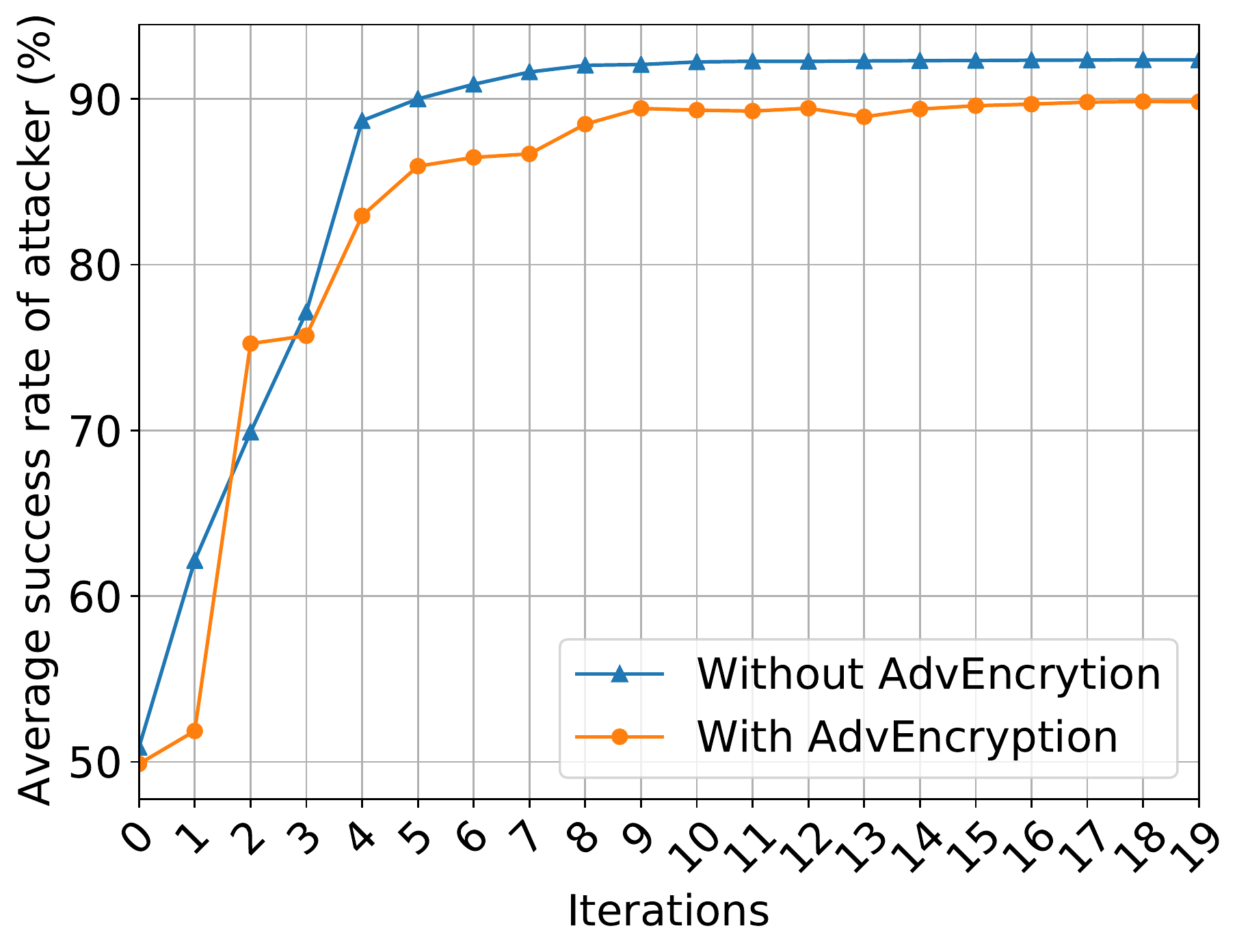}
         \caption{GAN-based method}
        \label{fig:dast_cicddos}
    \end{subfigure}
    \caption{Evaluation of our proposed AdvEncryption method using CICDDoS-2019 dataset from the prospective of average success rate of attacker. The attacker is assumed to construct the substitute model via: (a) Jacobian-based method, (a) FGSM-based method, (c) FGV-based method, and (d) GAN-based methods.}
    \label{fig:cicddos_comparison_iterations}
\end{figure}

\begin{figure}[!t]
    \centering
    \begin{subfigure}[t]{0.3\textwidth}
        \centering
        \includegraphics[width=\textwidth]{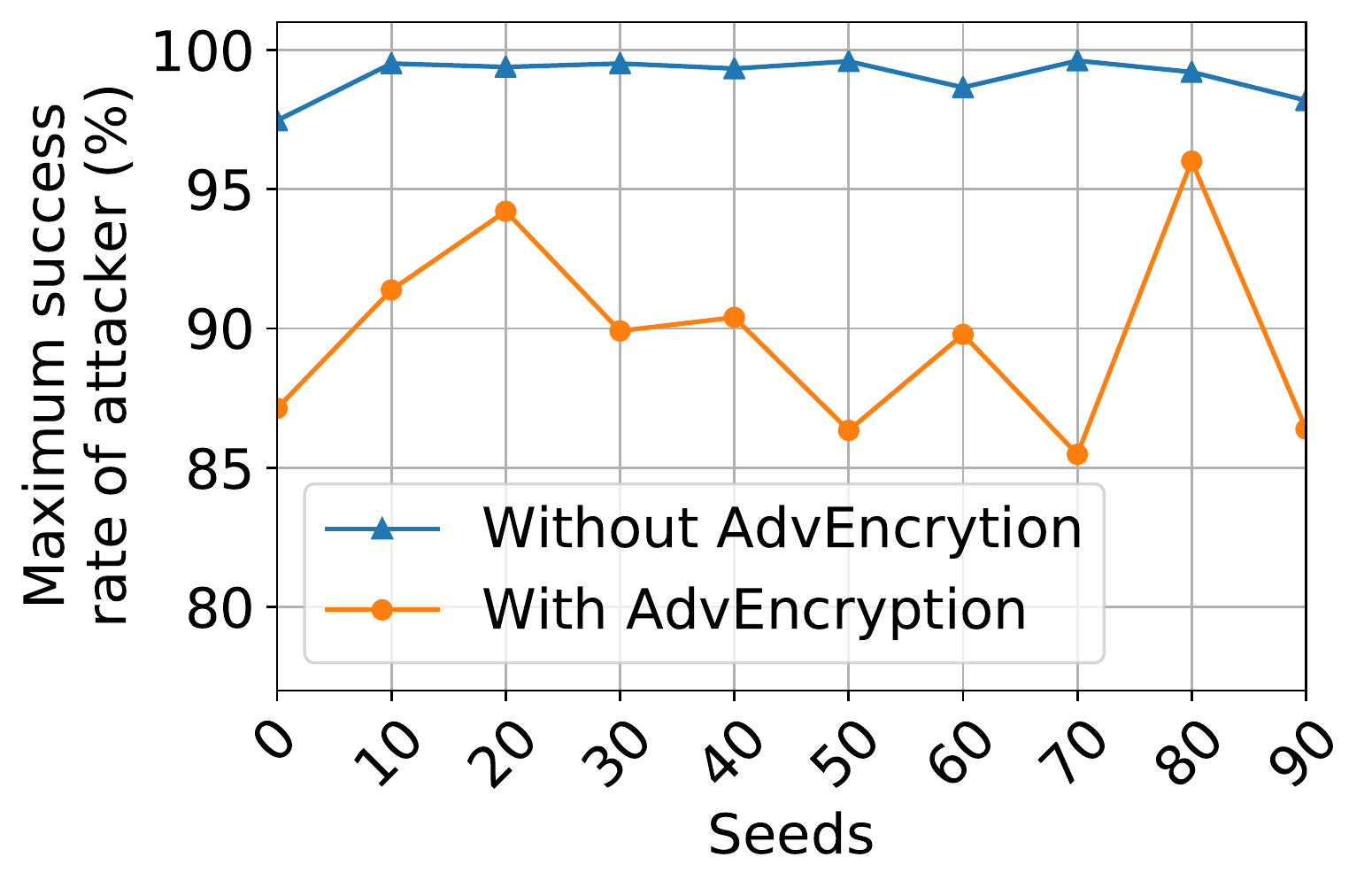}
        \caption{Jacobian-based method}
        \label{fig:jac_cicddos_seeds}
    \end{subfigure}
    \hfill
    \begin{subfigure}[t]{0.3\textwidth}
        \centering
        \includegraphics[width=\textwidth]{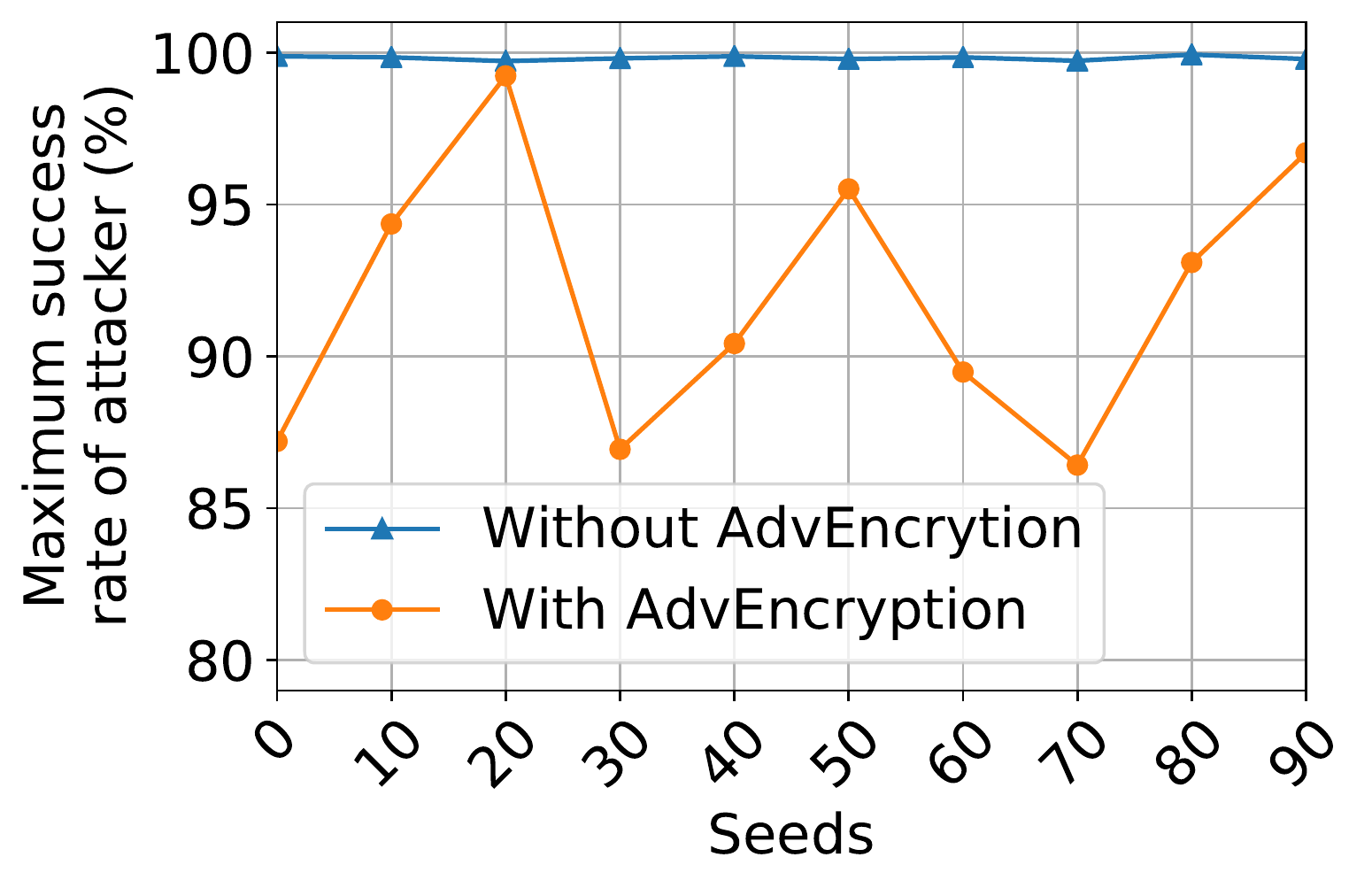}
        \caption{FGSM-based method}
        \label{fig:fgsm_cicddos_seeds}
    \end{subfigure}
    \hfill
    \begin{subfigure}[t]{0.3\textwidth}
        \centering
        \includegraphics[width=\textwidth]{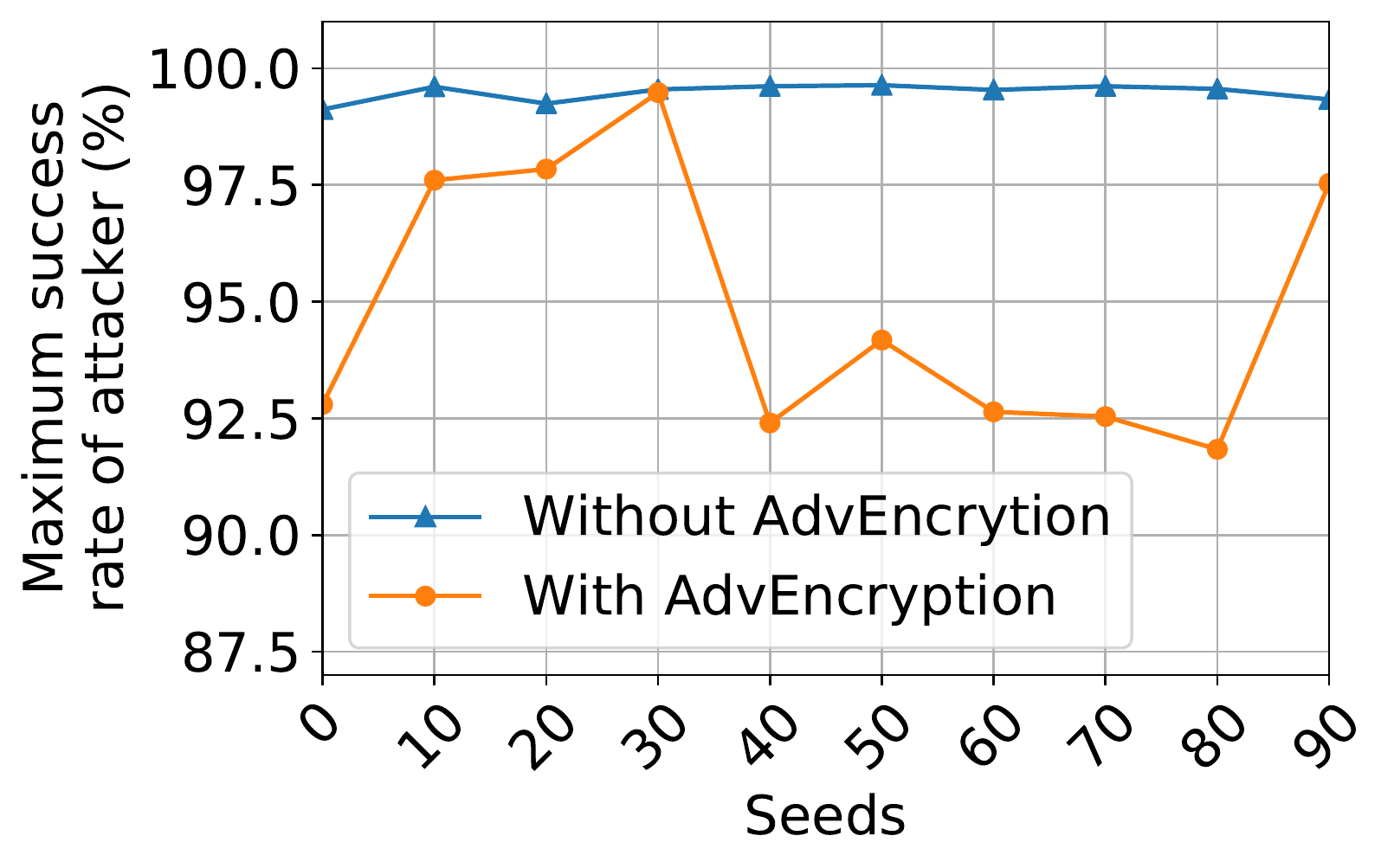}
        \caption{FGV-based method}
        \label{fig:fgv_cicddos_seeds}
    \end{subfigure}
    \caption{Evaluation of our proposed AdvEncryption method using CICDDoS-2019 dataset from the prospective of maximum success rate of attacker. The attacker is assumed to construct the substitute model via: (a) Jacobian-based method, (b) FGSM-based method, and (c) FGV-based methods.}
    \label{fig:cicddos_comparison_seeds}
\end{figure}

We continue to evaluate the performance of our proposed AdvEncryption method by measuring the success rate of the attacker on constructing the substitute model based on the feature distillation on the data streams of the targeted NN-based classifier. As shown in Fig. \ref{fig:jac_cicddos}, the attacker is assumed to construct the substitute model by using the Jacobian-based gradient method. The average success rate that the attacker can achieve, when the proposed AdvEncryption is integrated with the NN-based classifier, is shown in the orange plot, and the average success rate without AdvEncryption is shown in the blue plot. By comparing the success rates obtained with and without AdvEncryption, it is clear that our proposed AdvEncryption method can effectively reduce the success rate of the attacker. In other words, our proposed AdvEncryption method can effectively mislead the attacker and mitigate its capability of feature distillation on the data being used for autonomous cyber-attack detection. We also consider the situation when the attacker leverages FGSM-based, FGV-based, and GAN-based methods, respectively. In each situation, the average success rates with and without the AdvEncryption method are measured and shown in Figs. \ref{fig:fgsm_cicddos}, \ref{fig:fgv_cicddos}, and \ref{fig:dast_cicddos}, respectively. In Fig. \ref{fig:cicddos_comparison_iterations}, we can observe that, by deploying our proposed AdvEncryption, the success rate of the attacker is reduced significantly when the number of iterations is larger than 2. 

 Furthermore, the maximum success rates of the attacker for different simulation seeds are plotted in Fig. \ref{fig:cicddos_comparison_seeds} where the attacher is assumed to use Jacobian-based, FGSM-based, and FGV-based methods, respectively. From Fig. \ref{fig:cicddos_comparison_seeds}, we can get that the maximum success rate of the attacker by implementing our AdvEncryption method is lower in the majority of the seeds compared to the situation when the AdvEncrytion is not implemented.

\subsection{Scenario 2: Handwritten Digit Recognition}

\subsubsection{Training details} 
\begin{figure}[!t]
    \centering
    \begin{subfigure}[t]{0.6\textwidth}
        \centering
        \includegraphics[width=\textwidth]{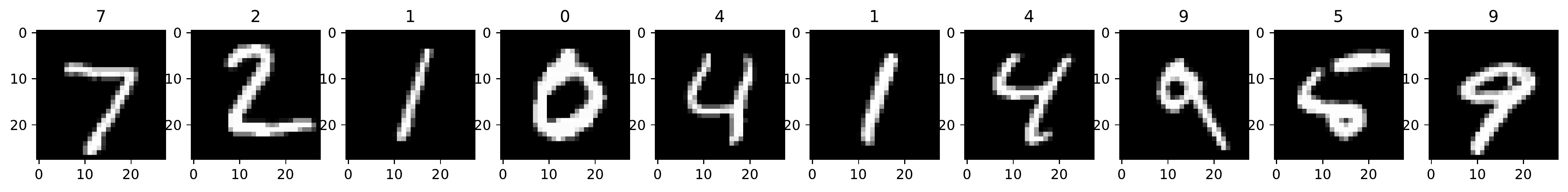}
        \caption{Sample of original MNIST images (i.e. plaintext data)}
        \label{fig:mnist_ori_images}
    \end{subfigure}
    \hfill
    \begin{subfigure}[t]{0.6\textwidth}
        \centering
        \includegraphics[width=\textwidth]{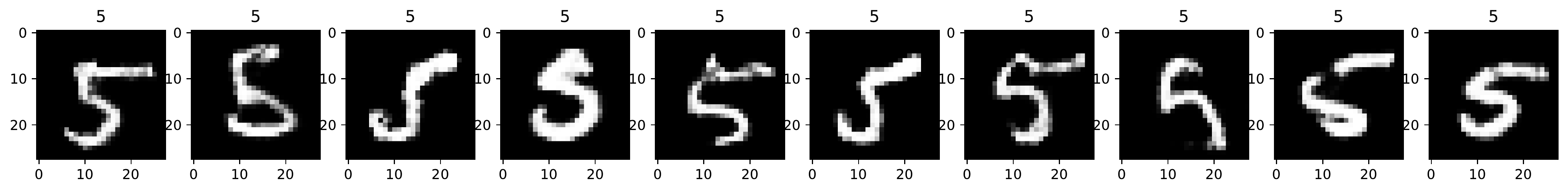}
        \caption{Encrypted MNIST images}
        \label{fig:mnist_enc_images}
    \end{subfigure}
    \hfill
    \begin{subfigure}[t]{0.6\textwidth}
        \centering
        \includegraphics[width=\textwidth]{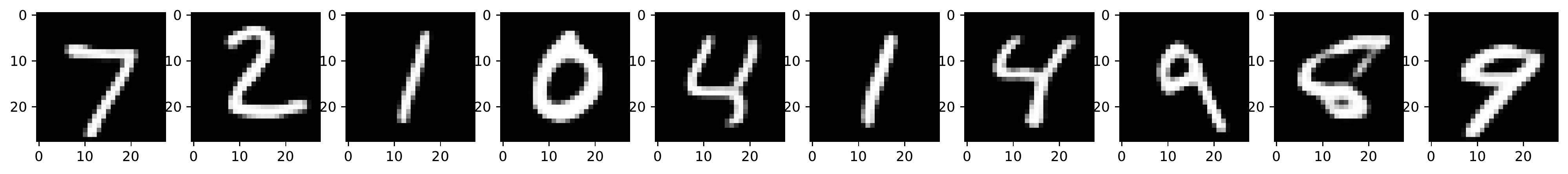}
        \caption{Decrypted MNIST images}
        \label{fig:mnist_dec_images}
    \end{subfigure}
    \caption{Example of plaintext (a), encrypted (b), and decrypted (c) MNIST images.}
    \label{fig:mnist_images}
\end{figure}

In this scenario, we use MNIST dataset \cite{LeCun2010MNISTDatabase} to evaluate the performance of our proposed method. Additionally, we use convolutional neural networks (CNN) to realize the AdvEncryption method and the baseline classifier. The training and testing accuracies of the CNN-based classifier with and without using our proposed AdvEncryption, respectively, are shown in Table \ref{tab:mnist_acc}. From Table \ref{tab:mnist_acc}, it can be observed that, by integrating AdvEncryption, the accuracy of the CNN-based classifier is slightly reduced. This is reasonable, due to the additional noise introduced during encryption and decryption. However, the training and testing accuracies remain at an efficiently high level. Our proposed method can achieve the encryption and decryption efficiencies of over $97\%$. The encrypted images and their associated plaintext images are presented in Fig. \ref{fig:mnist_images}. It can be observed that, by using our method, the plaintext images with different labels are successfully encrypted to images that correspond to label $5$, which is the targeted fake label. 

\begin{table}[t]
  \caption{Training and testing accuracies (MNIST)}
  \label{tab:mnist_acc}
  \centering
  \begin{tabular}{p{0.2\linewidth} p{0.3\linewidth} p{0.3\linewidth}}
    \toprule
    & Training accuracy & Testing accuracy \\
    \midrule
    Without AdvEncryption & 99.73000\% & 99.22000\%\\
    With AdvEncryption    & 98.35500\% & 97.04000\%\\
    \midrule
   \multicolumn{3}{l}{Encryption efficiency : 98.01000\%}\\
    \midrule
    \multicolumn{3}{l}{Decryption efficiency : 97.26886\%}\\
    \bottomrule
  \end{tabular}
\end{table}

\subsubsection{Performance evaluation}

\begin{figure}[!t]
    \centering
    \begin{subfigure}[t]{0.3\textwidth}
        \centering
        \includegraphics[width=\textwidth]{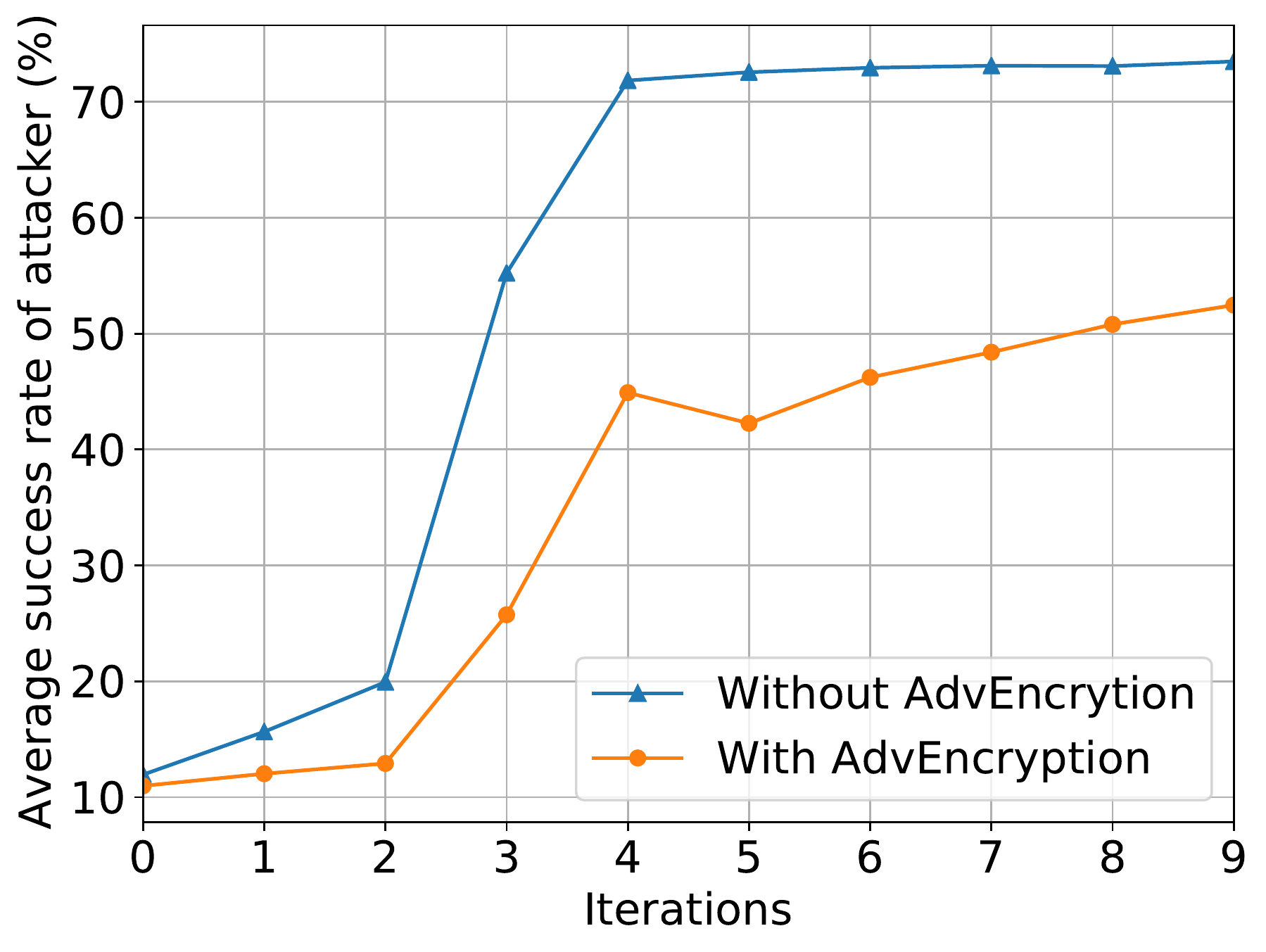}
        \caption{Jacobian-based method}
        \label{fig:jac_mnist}
    \end{subfigure}
    \hfill
    \begin{subfigure}[t]{0.3\textwidth}
        \centering
        \includegraphics[width=\textwidth]{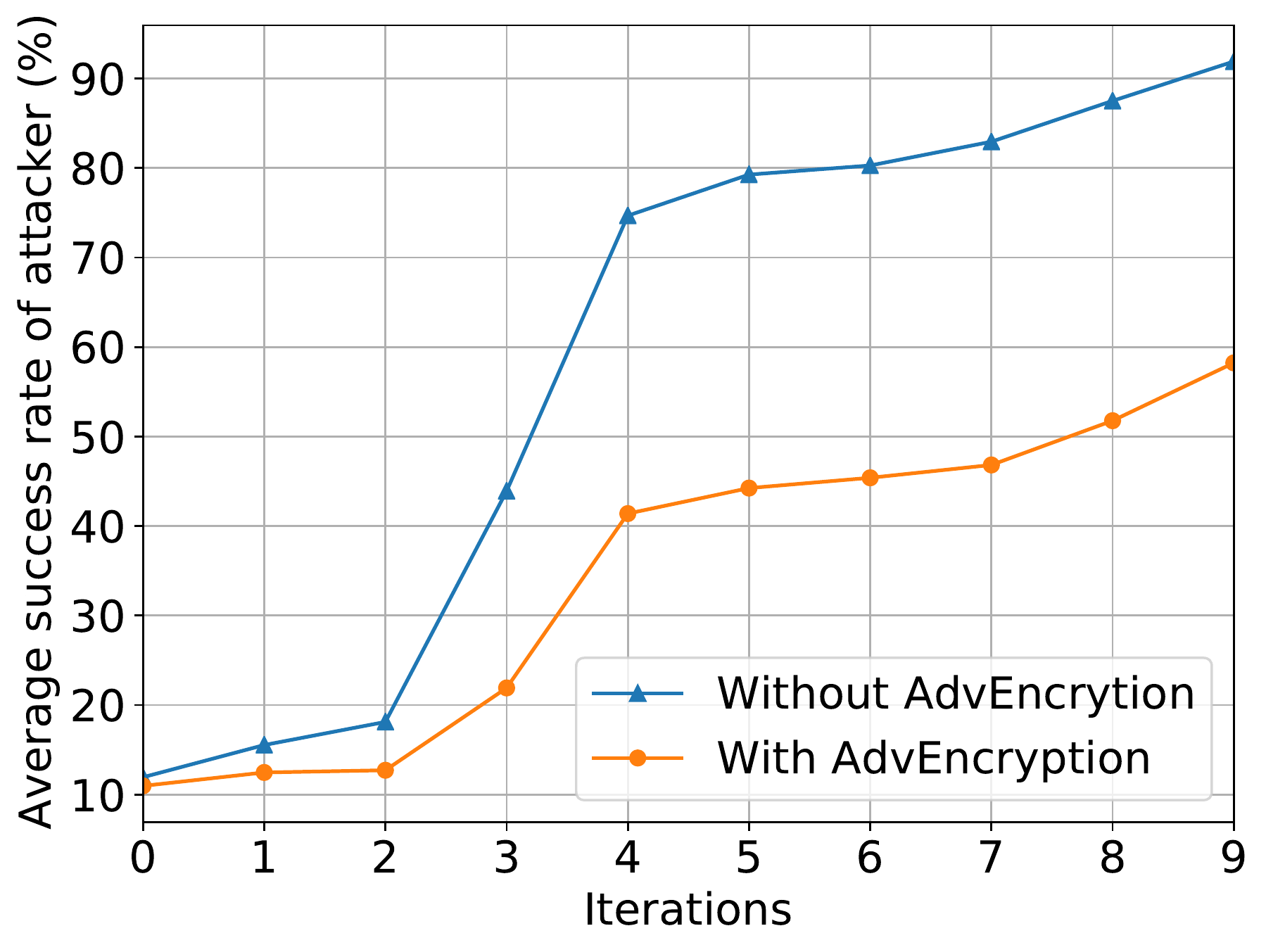}
        \caption{FGSM-based method}
        \label{fig:fgsm_mnist}
    \end{subfigure}
    \hfill
    \begin{subfigure}[t]{0.3\textwidth}
        \centering
        \includegraphics[width=\textwidth]{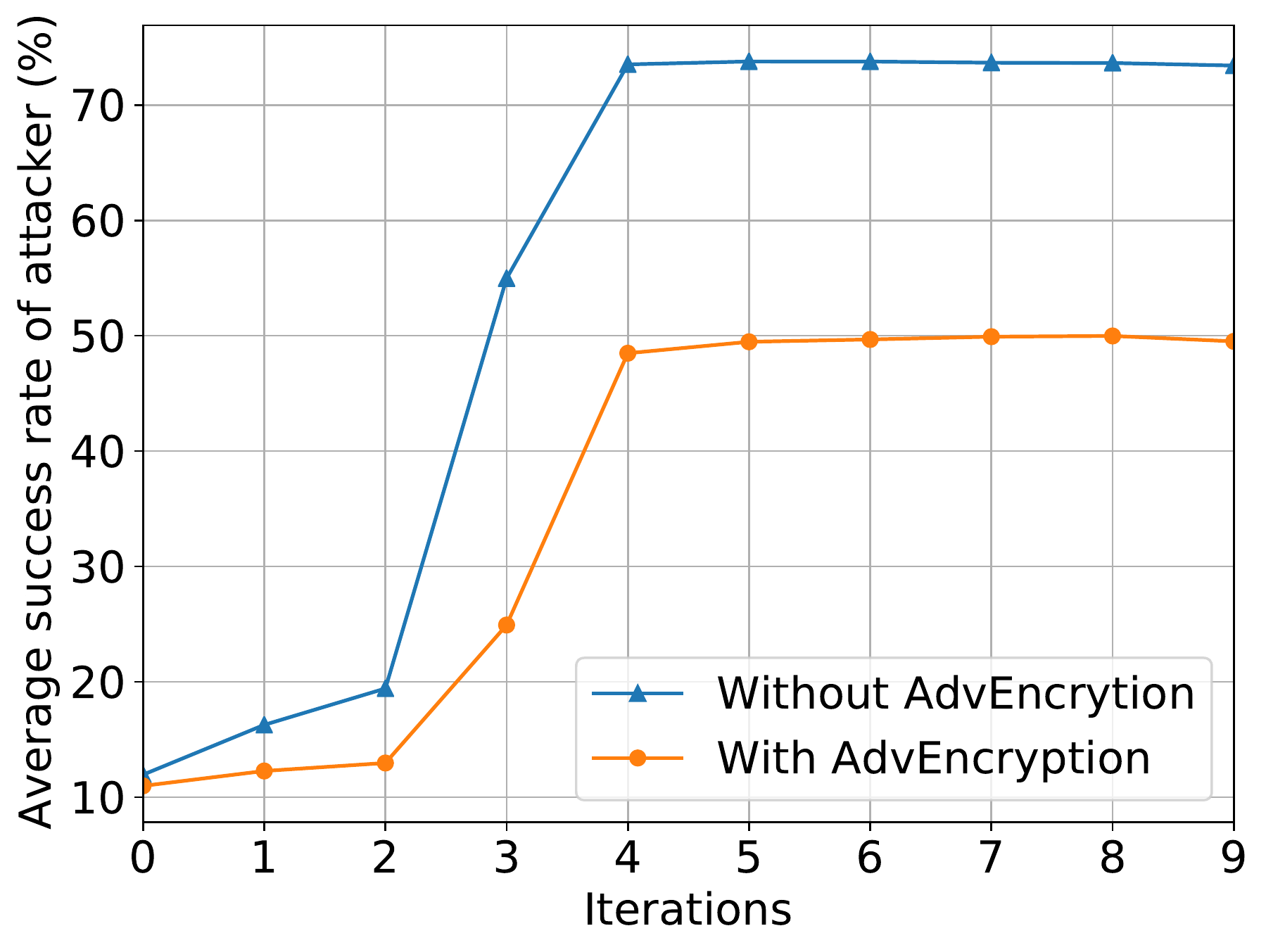}
        \caption{FGV-based method}
        \label{fig:fgv_mnist}
    \end{subfigure}
    \caption{Evaluation   of   our   proposed   AdvEncryption method using MNIST dataset from the prospective of average success rate of attacker. The attacker is assumed to  construct  the  substitute  model  via:  (a)  Jacobian-based method, (a) FGSM-based method, and (c) FGV-based methods.}
    \label{fig:mnist_comparison_iterations}
\end{figure}

\begin{figure}[!t]
    \centering
    \begin{subfigure}[b]{0.3\textwidth}
        \centering
        \includegraphics[width=\textwidth]{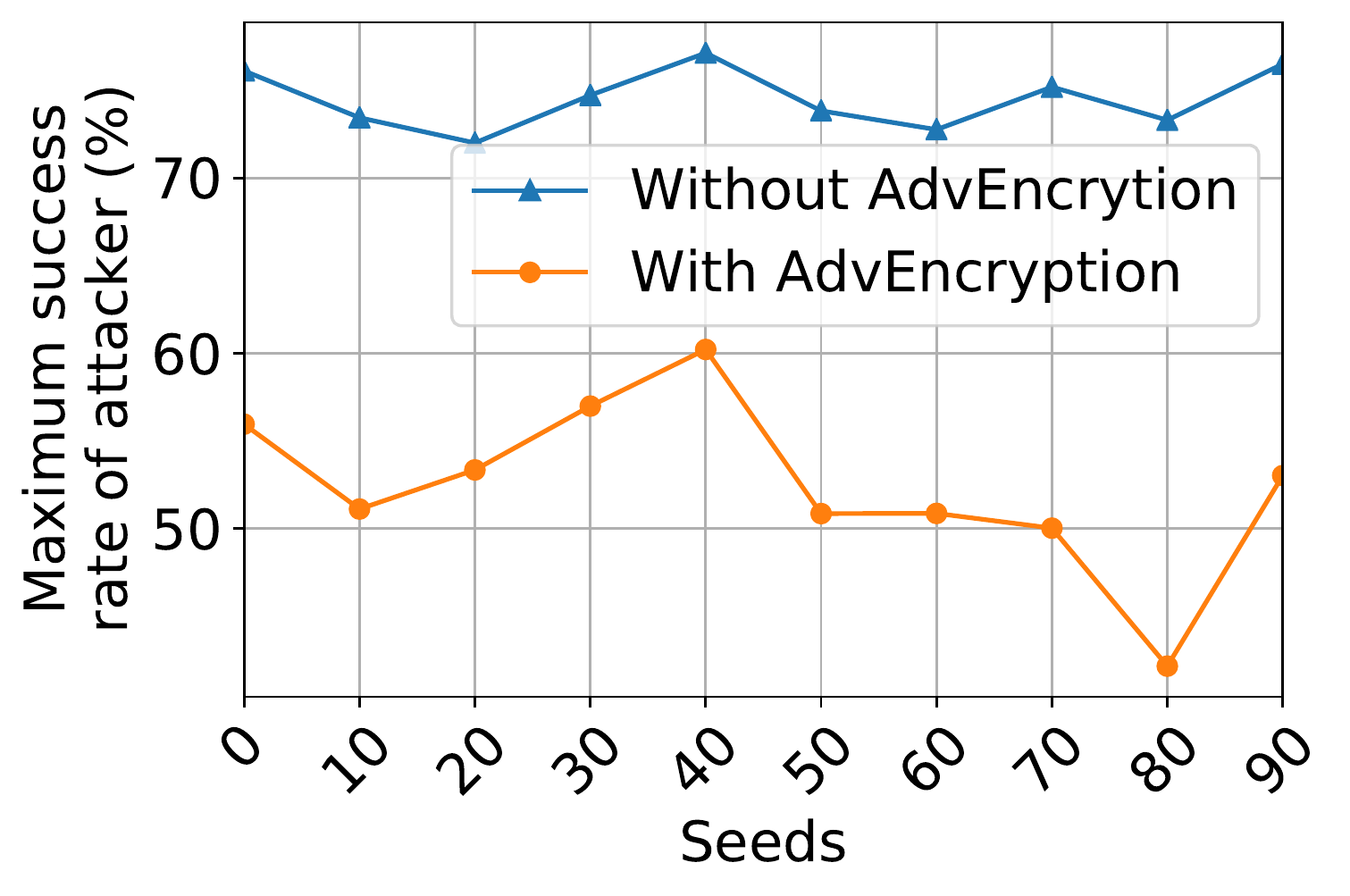}
        \caption{Jacobian-based method}
        \label{fig:jac_mnist_seeds}
    \end{subfigure}
    \hfill
    \begin{subfigure}[b]{0.3\textwidth}
        \centering
        \includegraphics[width=\textwidth]{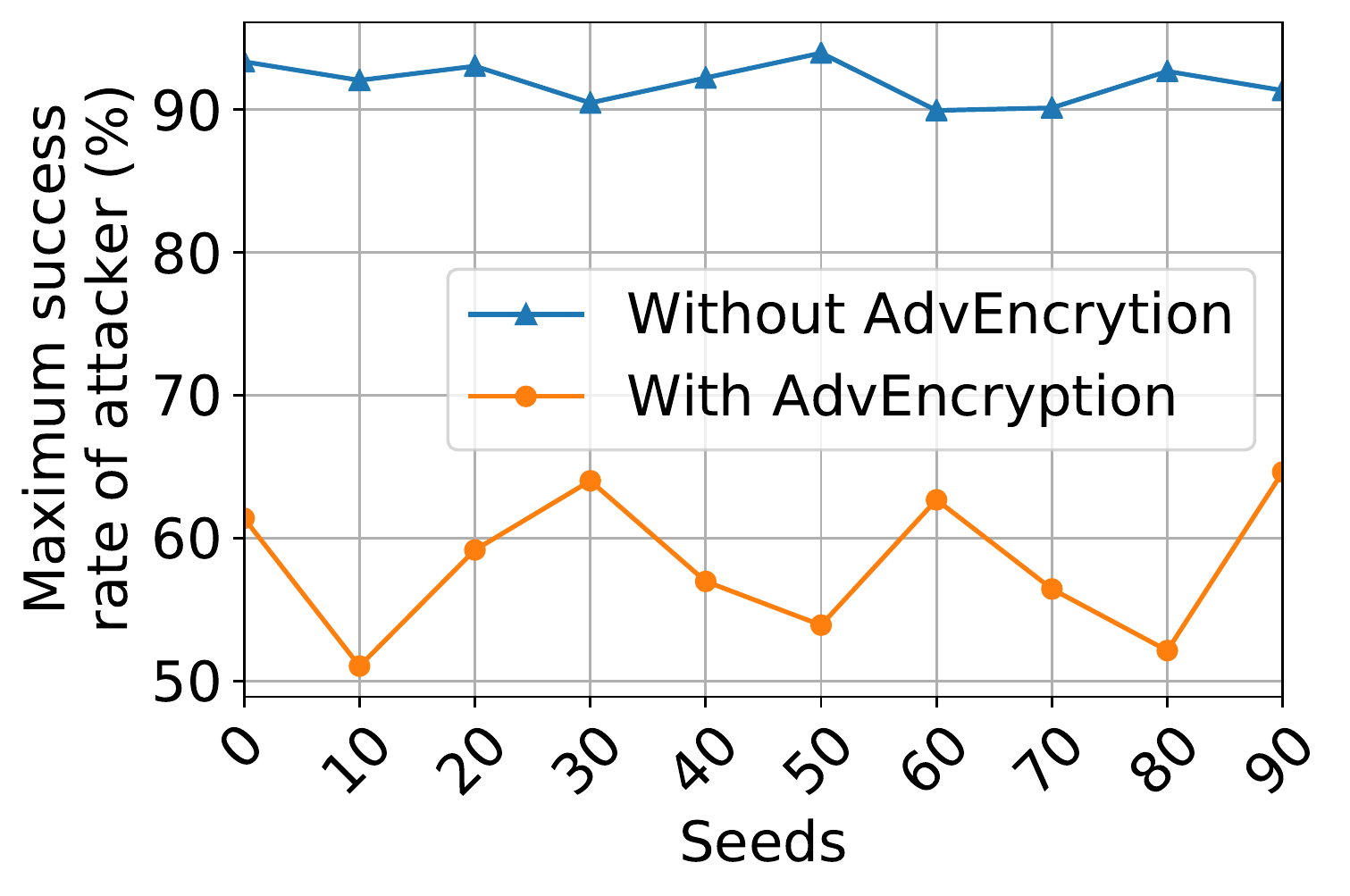}
        \caption{FGSM-based method}
        \label{fig:fgsm_mnist_seeds}
    \end{subfigure}
    \hfill
    \begin{subfigure}[b]{0.3\textwidth}
        \centering
        \includegraphics[width=\textwidth]{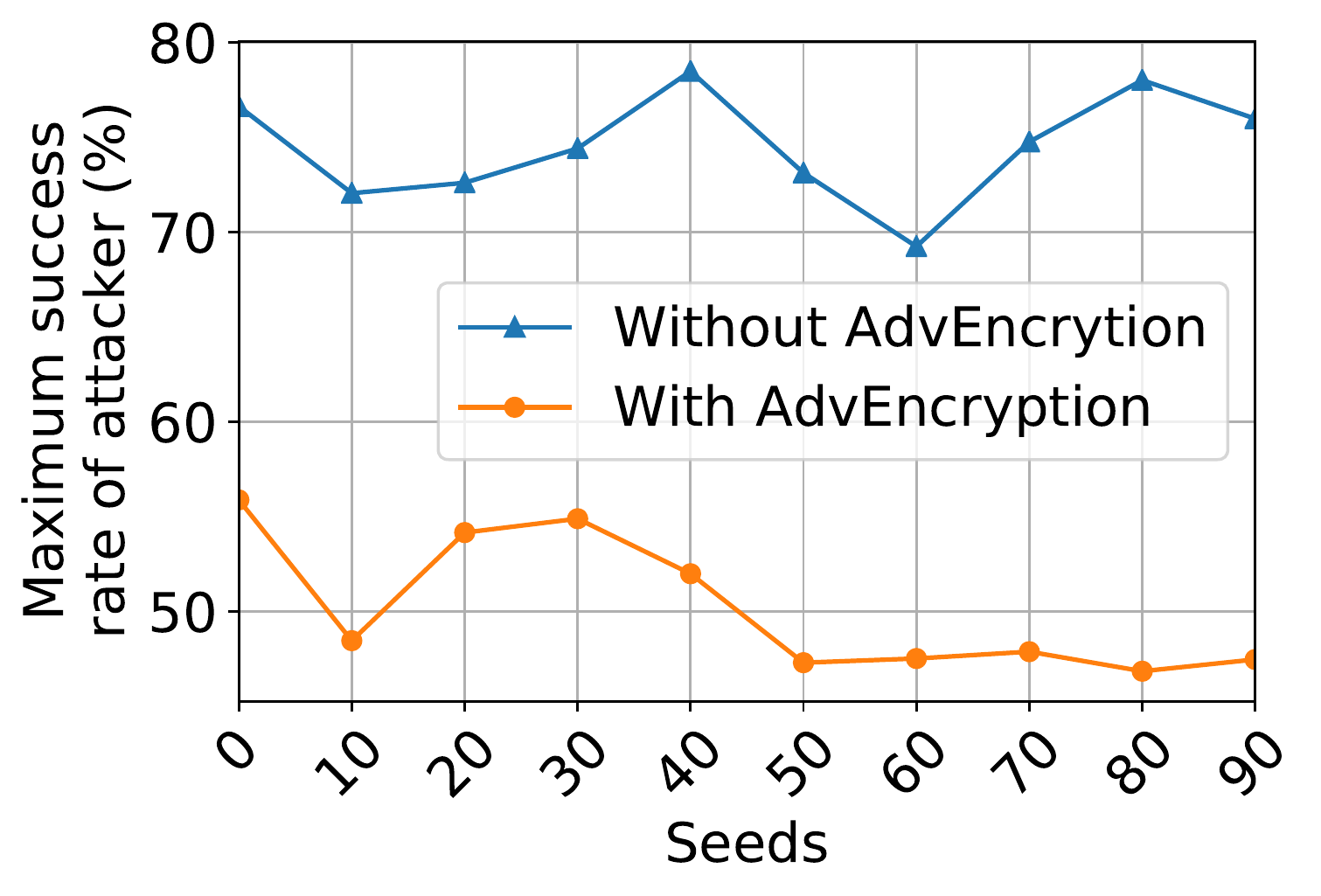}
        \caption{FGV-based method}
        \label{fig:fgv_mnist_seeds}
    \end{subfigure}
    \caption{Evaluation   of   our   proposed   AdvEncryption method using MNIST dataset from the prospective of  maximum  success  rate  of  attacker.  The  attacker  is  assumed to construct the substitute model via: (a) Jacobian-based method, (b) FGSM-based method, and (c) FGV-based methods.}
    \label{fig:mnist_comparison_seeds}
\end{figure}

Similar to Scenario 1, we evaluate the performance of our proposed AdvEncryption method by measuring the success rate of the attacker. As shown in Fig. \ref{fig:mnist_comparison_iterations}, the attacker is assumed to construct the substitute model by using Jacobian-based, FGSM-based, and FGV-based methods, respectively. The average success rate that the attacker can achieve, when the proposed AdvEncryption method is integrated with the CNN-based classifier, is shown in the orange plot. The average success rate without AdvEncryption is shown in the blue plot. Comparing the success rates obtained with and without AdvEncryption, it is clear that our proposed AdvEncryption method can effectively reduce the capability of the attacker on feature distillation on the data being used for autonomous handwritten digit recognition. Furthermore, the maximum success rates of the attacker for different simulation seeds are plotted in Fig. \ref{fig:mnist_comparison_seeds} where the attacker is assumed to use different gradient-based methods. From Fig. \ref{fig:mnist_comparison_seeds}, we can get that the maximum success rate of the attacker by implementing our AdvEncryption method is lower in all the seeds compared to the situation when the AdvEncrytion is not implemented.
 
\subsection{Scenario 3: Facial Recognition}

\subsubsection{Training details}

\begin{table*}[!b]
  	\caption{Training and testing accuracies (CelebA)}
 	 \label{tab:celeba_acc}
  	\centering
  	\begin{tabular}{p{0.3\linewidth} p{0.2\linewidth} p{0.15\linewidth} p{0.15\linewidth}}
    \toprule
    & Classification task & Training accuracy & Testing accuracy\\
 	\midrule
    Without proposed AdvEncryption & Gender recognition & 97.58002\% & 95.19086\%\\
    & Smiling recognition & 91.29569\% & 90.23645\%\\
   	\midrule
   	With proposed AdvEncryption & Gender recognition & 95.44142\% & 94.47951\%\\
    & Smiling recognition & 91.14272\% & 90.11622\%\\
    \midrule
    Encryption efficiency & Gender recognition &  92.49574\%\\
    &  Smiling recognition &   90.73239\%\\
    \midrule
    \multicolumn{1}{l}{Decryption efficiency : 94.72330\%}\\
    \bottomrule
  \end{tabular}
\end{table*}

In this scenario, we evaluate the performance of our proposed AdvEncryption method on simultaneously securing two features of CelebA dataset \cite{Liu2015DeepWild}, including smile (smiling/not smiling) and gender (male/female), against attackers. In other words, the encoder and decoder of the AdvEncryption are trained to mislead an attacker on both recognition tasks. To achieve this goal, given a feature in CelebA dataset, the encryption mechanism in our proposed method converts the data feature into the randomly-selected unified fake two-dimensional feature ((smiling, male) is selected as the fake feature in our case study). Additionally, we use CNNs to realize the AdvEncryption method and the baseline classifier.

The training and testing accuracies of the CNN-based classifier for both gender and smiling recognition tasks with and without using our proposed AdvEncryption, respectively, are shown in Table \ref{tab:celeba_acc}. From Table \ref{tab:celeba_acc}, it can be observed that, by integrating AdvEncryption, the accuracy of the CNN-based classifier is slightly reduced due to the additional noise introduced during the encryption and decryption of the input data. However, the accuracy reduction is very limited and the training and testing accuracies remain at an efficiently high level. Additionally, our proposed method can achieve encryption and decryption efficiencies above $90\%$ for both recognition tasks.  

\subsubsection{Performance evaluation}

\begin{figure}[!t]
    \centering
    \begin{subfigure}[t]{0.3\textwidth}
        \centering
        \includegraphics[width=\textwidth]{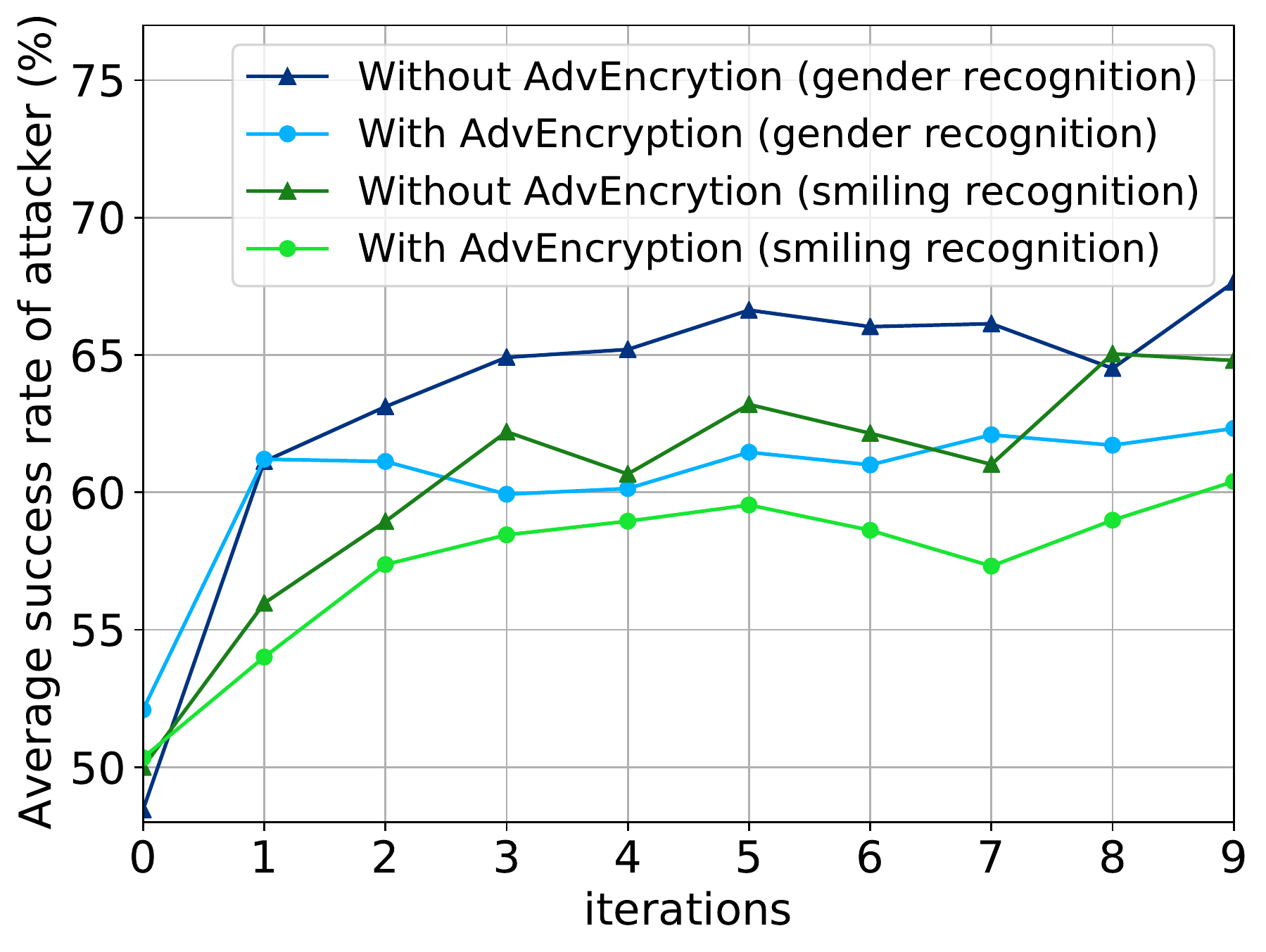}
        \caption{Jacobian-based method}
        \label{fig:jac_task_celeba}
    \end{subfigure}
    \hfill
    \begin{subfigure}[t]{0.3\textwidth}
        \centering
        \includegraphics[width=\textwidth]{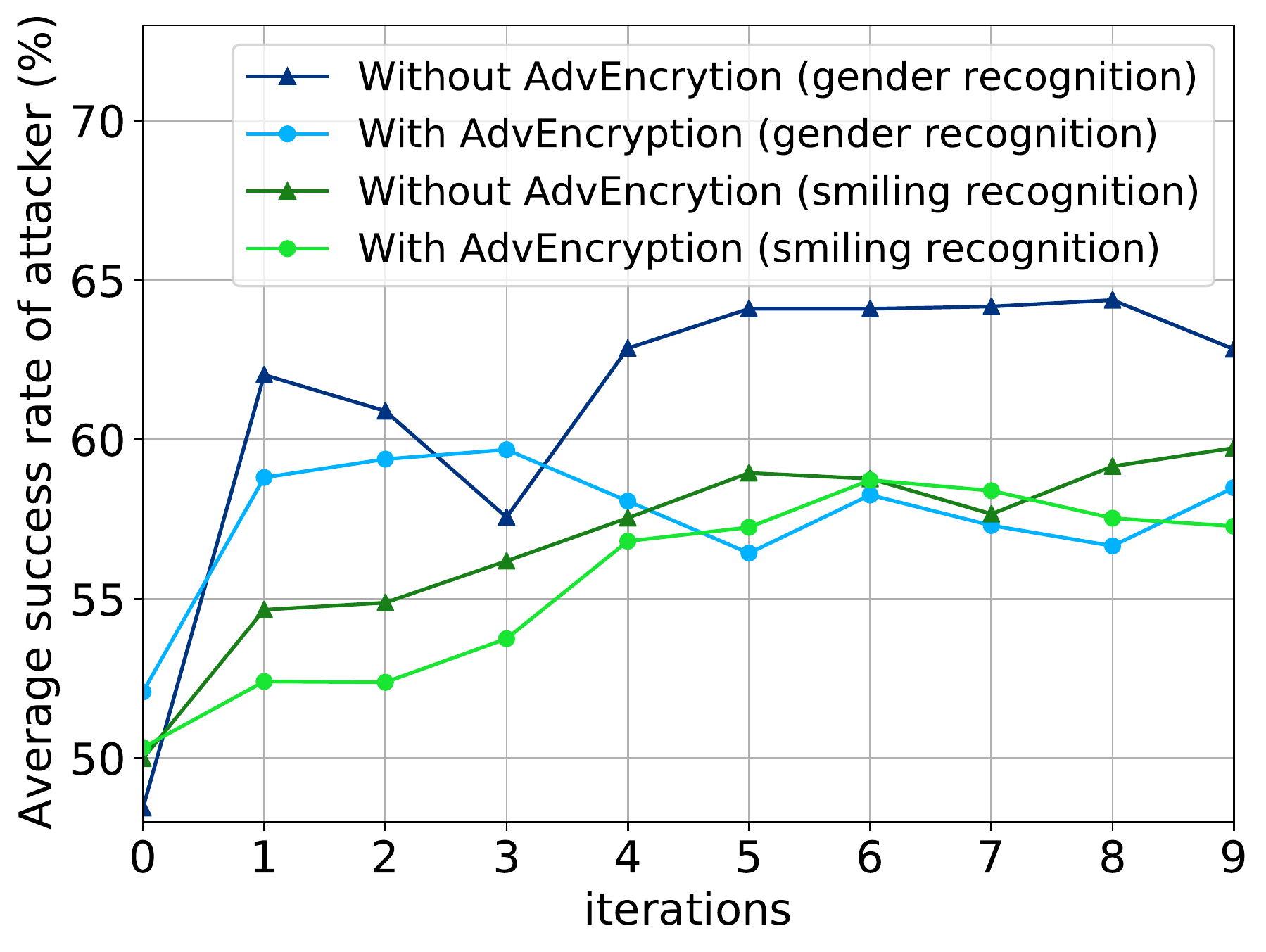}
        \caption{FGSM-based method}
        \label{fig:fgsm_task_celeba}
    \end{subfigure}
    \hfill
    \begin{subfigure}[t]{0.3\textwidth}
        \centering
        \includegraphics[width=\textwidth]{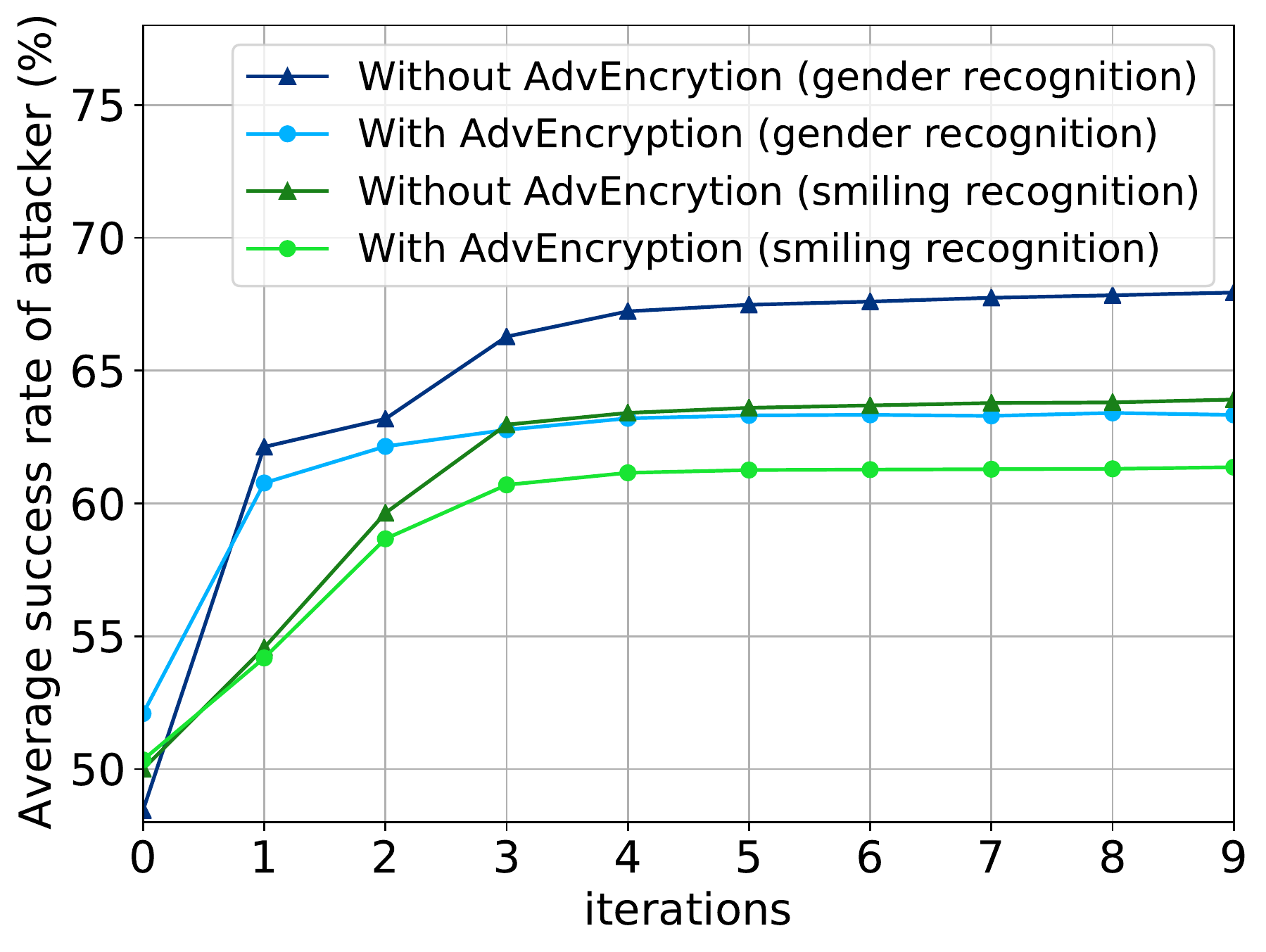}
        \caption{FGV-based method}
        \label{fig:fgv_task_celeba}
    \end{subfigure}
    \hfill
    \begin{subfigure}[t]{0.3\textwidth}
        \centering
        \includegraphics[width=\textwidth]{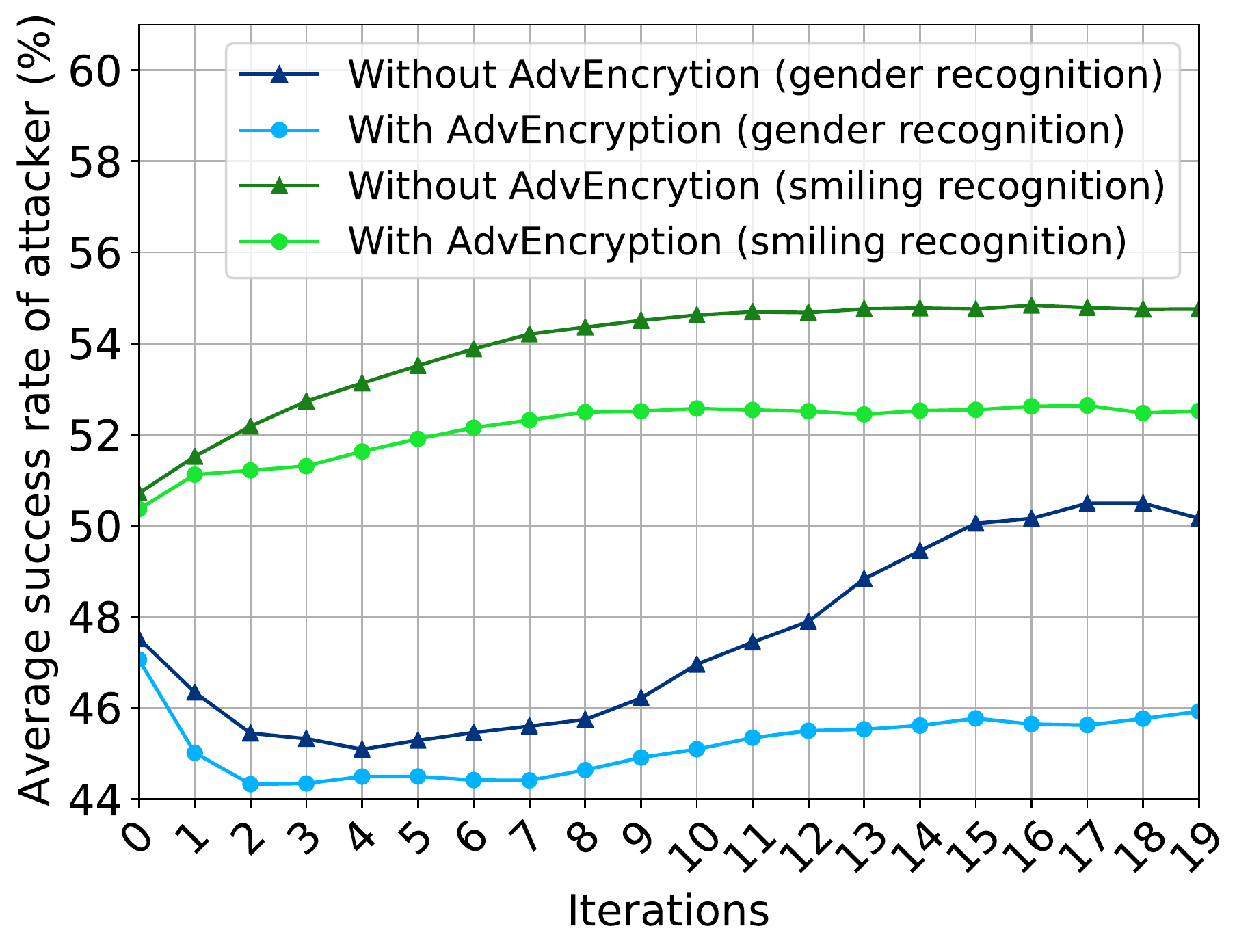}
         \caption{GAN-based attack}
        \label{fig:dast_task_celeba}
    \end{subfigure}
    \caption{Evaluation of our proposed AdvEncryption method using CelebA dataset from the prospective of average success rate of attacker. The attacker is assumed to construct the substitute model via: (a) Jacobian-based method, (a) FGSM-based method, (c) FGV-based, and (d) GAN-based methods.}
    \label{fig:celeba_task_comparison_iterations}
\end{figure}

\begin{figure}[!t]
    \centering
    \begin{subfigure}[t]{0.3\textwidth}
        \centering
        \includegraphics[width=\textwidth]{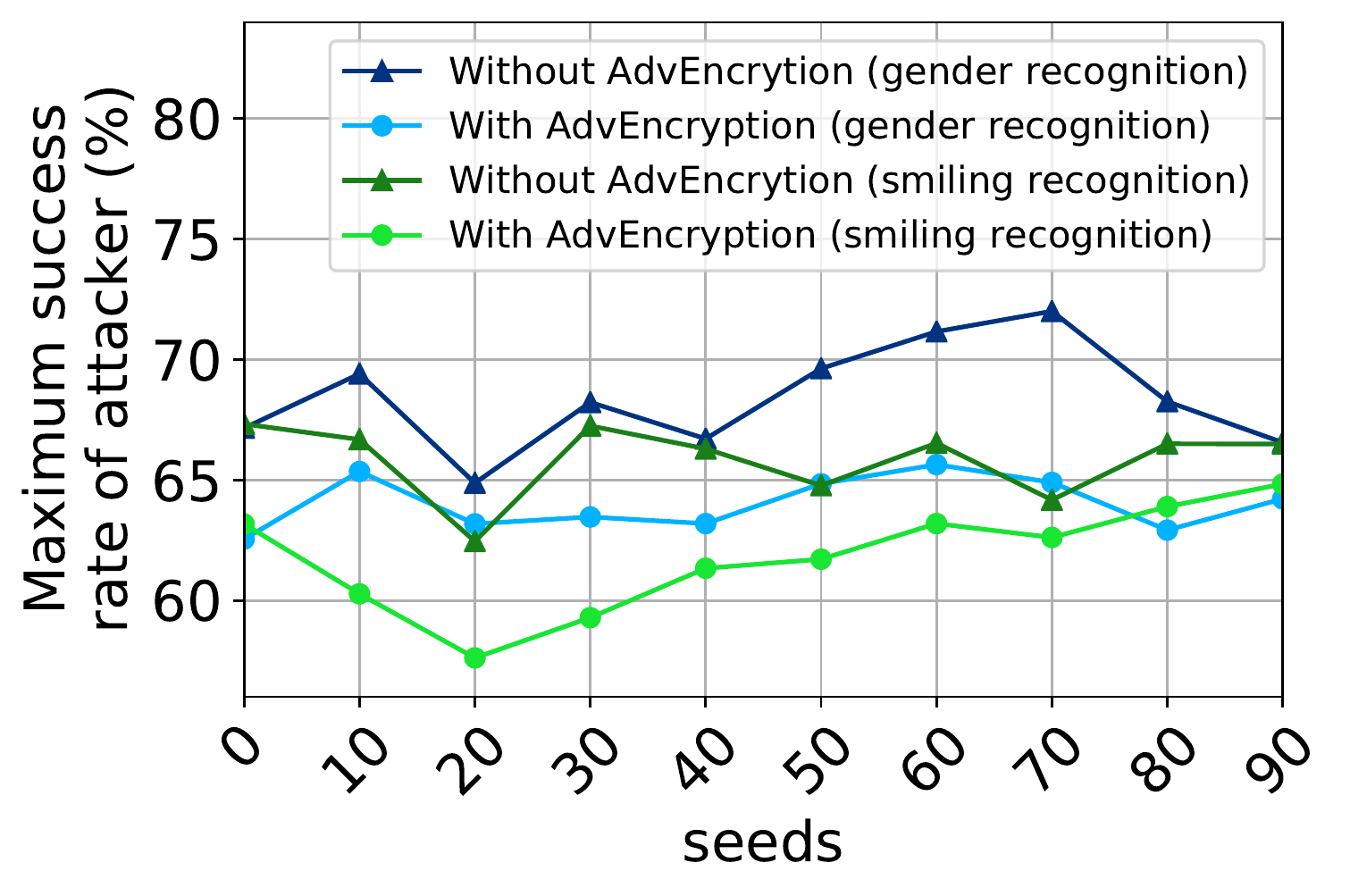}
        \caption{Jacobian-based method}
        \label{fig:jac_task_celeba_seeds}
    \end{subfigure}
    \hfill
    \begin{subfigure}[t]{0.3\textwidth}
        \centering
        \includegraphics[width=\textwidth]{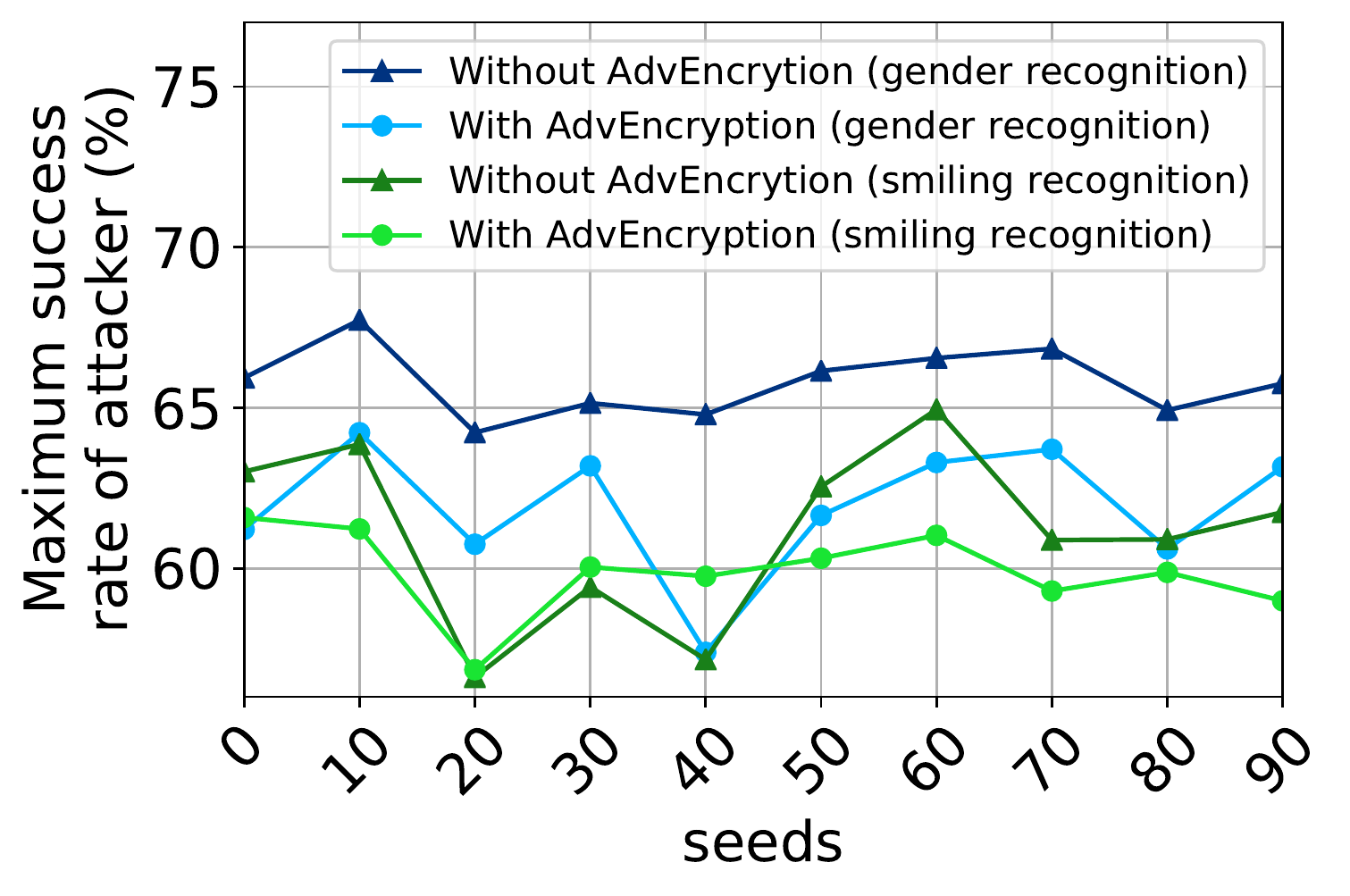}
        \caption{FGSM-based method}
        \label{fig:fgsm_task_celeba_seeds}
    \end{subfigure}
    \hfill
    \begin{subfigure}[t]{0.3\textwidth}
        \centering
        \includegraphics[width=\textwidth]{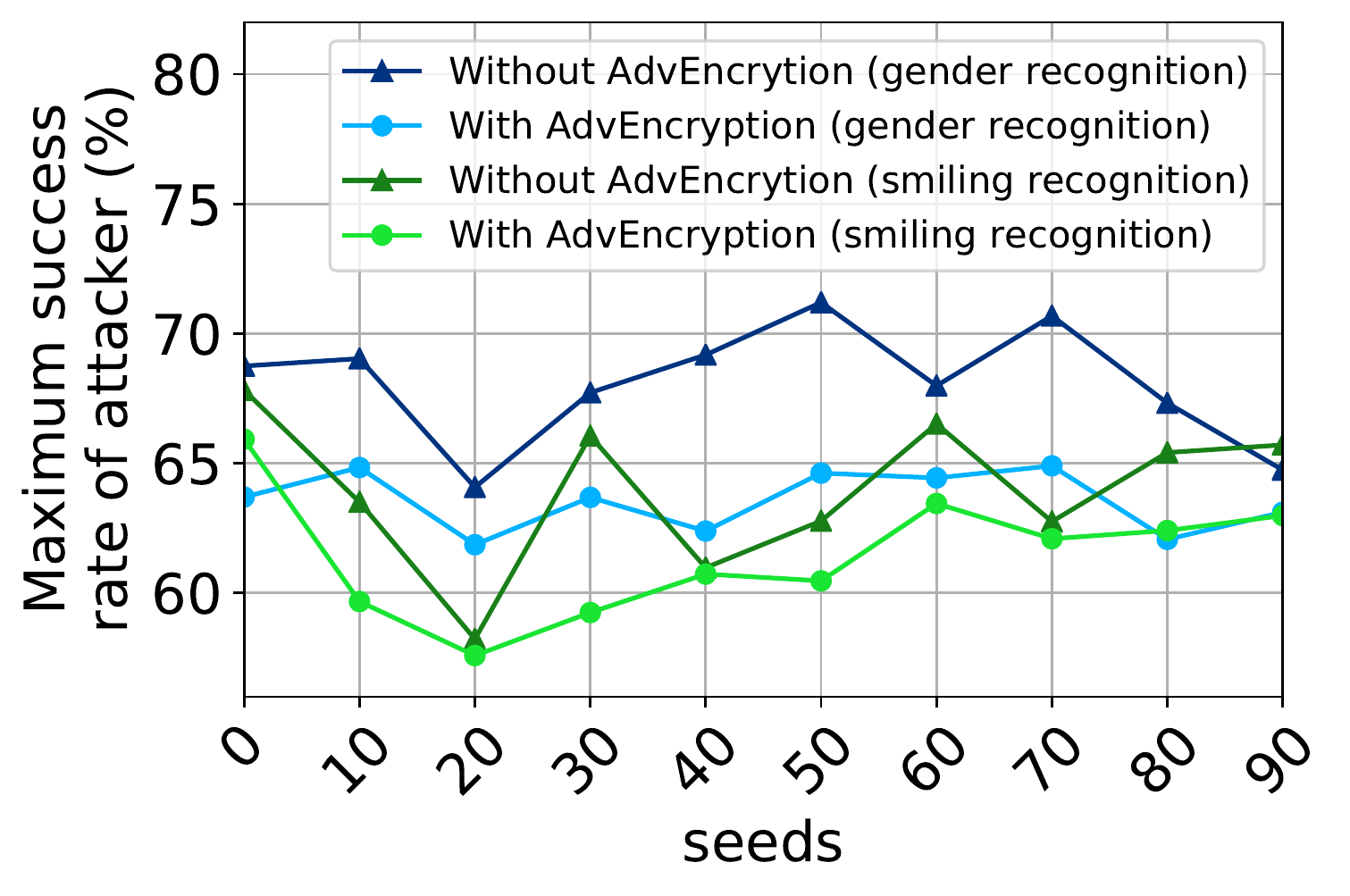}
        \caption{FGV-based method}
        \label{fig:fgv_task_celeba_seeds}
    \end{subfigure}
    \caption{Evaluation of our proposed AdvEncryption method using CelebA dataset from the prospective of maximum success rate of attacker. The attacker is assumed to construct the substitute model via: (a) Jacobian-based method, (a) FGSM-based method, and (c) FGV-based methods.}
    \label{fig:celeba_task_comparison_seeds}
\end{figure}

In this case study, we evaluate the performance in gender recognition and smiling recognition tasks individually. As shown in Fig. \ref{fig:jac_task_celeba}, the attacker is assumed to use the Jacobian-based gradient method, to construct the substitute model to emulate the behavior of CNN-based classifier for the gender and smiling recognition tasks. For the gender recognition task, the average success rate that the attacker can achieve, when the proposed AdvEncryption is integrated with the CNN-based classifier, is shown in the light blue plot and the average success rate without AdvEncryption is shown in the dark blue plot. Comparing these two plots, it is clear that our proposed AdvEncryption method can effectively reduce the success rate of the attacker. In other words, our proposed AdvEncryption method can effectively mislead the attacker and mitigate its capability of feature distillation on the data being used for autonomous decision making on gender recognition. A similar result can be observed for smiling recognition in Fig. \ref{fig:jac_task_celeba}. We also consider the situation when the attacker leverages FGSM-based, FGV-based gradient methods and GAN-based methods, respectively, for both gender and smiling recognition tasks. In each situation, the average success rates with and without the AdvEncryption method are measured and shown in Figs. \ref{fig:fgsm_task_celeba} to \ref{fig:dast_task_celeba}, respectively. From Figs. \ref{fig:fgsm_task_celeba} to \ref{fig:dast_task_celeba}, we can observe that, by deploying our proposed AdvEncryption, the success rate of the attacker stays low throughout the majority of the iterations.

We also evaluate the maximum success rates of the attacker for different simulation seeds and present the simulation results in Figs. \ref{fig:jac_task_celeba_seeds} to \ref{fig:fgv_task_celeba_seeds}, where the attacher is assumed to use Jacobian-based, FGSM-based, and FGV-based methods, respectively. The plots represent both smiling and gender recognition tasks. From the plots, we can observe that the maximum success rate of the attacker when the AdvEncryption is implemented is significantly lower in a majority of the seeds compared with the situation when the proposed AdvEncryption is not implemented.

\section{Conclusions}

In this paper, we develop an innovative encryption method, AdvEncryption, by exploiting adversarial attack techniques. Different from existing encryption technologies that normally aim to prevent attackers from exploiting the dataset and can result in attack-vector changes on the autonomous decision-making procedure, our proposed method aims to trick the attackers to exhaust their effort in a misleading feature distillation of the data. To achieve this goal, our AdvEncryption method consists of two essential components: 1) an adversarial attack-inspired encryption mechanism that encrypts the data with stealthy adversarial perturbations to trick the attackers to consider encrypted data as plaintext data, and 2) a decryption mechanism that minimizes interruption on the autonomous decision-making procedure caused by the perturbations. As illustrated in our performance evaluation section, our proposed method is able to effectively mitigate the attackers' capability on feature distillation on the dataset while minimizing the impact of the perturbations on the effectiveness of autonomous decision making. In our ongoing work, we are evaluating our proposed method in more scenarios with different levels of scalability and complexity. 

\bibliographystyle{IEEEtran}
\bibliography{IEEEabrv, bare_conf}

\begin{thebibliography}{10}
\providecommand{\url}[1]{#1}
\csname url@samestyle\endcsname
\providecommand{\newblock}{\relax}
\providecommand{\bibinfo}[2]{#2}
\providecommand{\BIBentrySTDinterwordspacing}{\spaceskip=0pt\relax}
\providecommand{\BIBentryALTinterwordstretchfactor}{4}
\providecommand{\BIBentryALTinterwordspacing}{\spaceskip=\fontdimen2\font plus
\BIBentryALTinterwordstretchfactor\fontdimen3\font minus
  \fontdimen4\font\relax}
\providecommand{\BIBforeignlanguage}[2]{{%
\expandafter\ifx\csname l@#1\endcsname\relax
\typeout{** WARNING: IEEEtran.bst: No hyphenation pattern has been}%
\typeout{** loaded for the language `#1'. Using the pattern for}%
\typeout{** the default language instead.}%
\else
\language=\csname l@#1\endcsname
\fi
#2}}
\providecommand{\BIBdecl}{\relax}
\BIBdecl

\bibitem{Grigorescu2020ADriving}
S.~Grigorescu, B.~Trasnea, T.~Cocias, and G.~Macesanu, ``{A survey of deep
  learning techniques for autonomous driving},'' \emph{Journal of Field
  Robotics}, vol.~37, no.~3, 4 2020.

\bibitem{Devlin2019BERT:Understanding}
J.~Devlin, M.-W. Chang, K.~Lee, and K.~Toutanova, ``{BERT: Pre-training of Deep
  Bidirectional Transformers for Language Understanding},'' in \emph{NAACL},
  2019.

\bibitem{Levine2018LearningCollection}
S.~Levine, P.~Pastor, A.~Krizhevsky, J.~Ibarz, and D.~Quillen, ``{Learning
  hand-eye coordination for robotic grasping with deep learning and large-scale
  data collection},'' \emph{The International Journal of Robotics Research},
  vol.~37, no. 4-5, 4 2018.

\bibitem{Carlini2017TowardsNetworks}
N.~Carlini and D.~Wagner, ``{Towards Evaluating the Robustness of Neural
  Networks},'' in \emph{2017 IEEE Symposium on Security and Privacy
  (SP)}.\hskip 1em plus 0.5em minus 0.4em\relax IEEE, 5 2017.

\bibitem{Carlini2018AudioSpeech-to-Text}
------, ``{Audio Adversarial Examples: Targeted Attacks on Speech-to-Text},''
  in \emph{2018 IEEE Security and Privacy Workshops (SPW)}.\hskip 1em plus
  0.5em minus 0.4em\relax San Francisco, CA, USA: IEEE, 5 2018, pp. 1--7.

\bibitem{Kumar2006FundamentalsCryptography}
\BIBentryALTinterwordspacing
S.~Kumar and T.~Wollinger, ``{Fundamentals of Symmetric Cryptography},'' in
  \emph{Embedded Security in Cars: Securing Current and Future Automotive IT
  Applications}, K.~Lemke, C.~Paar, and M.~Wolf, Eds.\hskip 1em plus 0.5em
  minus 0.4em\relax Berlin, Heidelberg: Springer Berlin Heidelberg, 2006, pp.
  125--143. [Online]. Available: \url{https://doi.org/10.1007/3-540-28428-1_8}
\BIBentrySTDinterwordspacing

\bibitem{Daemen2001SpecificationAES}
\BIBentryALTinterwordspacing
J.~Daemen and V.~Rijmen, ``{Specification for the Advanced Encryption Standard
  (AES)},'' Federal Information Processing Standards Publication 197, 2001.
  [Online]. Available:
  \url{http://csrc.nist.gov/publications/fips/fips197/fips-197.pdf}
\BIBentrySTDinterwordspacing

\bibitem{Koblitz2004ACryptosystems}
N.~Koblitz and A.~J. Menezes, ``{A Survey of Public-Key Cryptosystems},''
  \emph{SIAM Review}, vol.~46, no.~4, 1 2004.

\bibitem{Rivest1978ACryptosystems}
\BIBentryALTinterwordspacing
R.~L. Rivest, A.~Shamir, and L.~Adleman, ``{A Method for Obtaining Digital
  Signatures and Public-Key Cryptosystems},'' \emph{Commun. ACM}, vol.~21,
  no.~2, pp. 120--126, 2 1978. [Online]. Available:
  \url{https://doi.org/10.1145/359340.359342}
\BIBentrySTDinterwordspacing

\bibitem{Vizitiu2020ApplyingData}
A.~Vizitiu, C.~I. Nita, A.~Puiu, C.~Suciu, and L.~M. Itu, ``{Applying Deep
  Neural Networks over Homomorphic Encrypted Medical Data},''
  \emph{Computational and Mathematical Methods in Medicine}, vol. 2020, 4 2020.

\bibitem{Gilad-Bachrach2016CryptoNets:Accuracy}
\BIBentryALTinterwordspacing
R.~Gilad-Bachrach, N.~Dowlin, K.~Laine, K.~Lauter, M.~Naehrig, and J.~Wernsing,
  ``{CryptoNets: Applying Neural Networks to Encrypted Data with High
  Throughput and Accuracy},'' in \emph{Proceedings of The 33rd International
  Conference on Machine Learning}, ser. Proceedings of Machine Learning
  Research, M.~F. Balcan and K.~Q. Weinberger, Eds., vol.~48.\hskip 1em plus
  0.5em minus 0.4em\relax New York, New York, USA: PMLR, 9 2016, pp. 201--210.
  [Online]. Available:
  \url{https://proceedings.mlr.press/v48/gilad-bachrach16.html}
\BIBentrySTDinterwordspacing

\bibitem{Hesamifard2017CryptoDL:Data}
E.~Hesamifard, H.~Takabi, and M.~Ghasemi, ``{CryptoDL: Deep Neural Networks
  over Encrypted Data},'' \emph{ArXiv}, vol. abs/1711.05189, 2017.

\bibitem{Goodfellow2014GenerativeNetworks}
\BIBentryALTinterwordspacing
I.~J. Goodfellow, J.~Pouget-Abadie, M.~Mirza, B.~Xu, D.~Warde-Farley, S.~Ozair,
  A.~Courville, and Y.~Bengio, ``{Generative Adversarial Networks},''
  \emph{Communications of the ACM}, vol.~63, no.~11, pp. 139--144, 6 2014.
  [Online]. Available: \url{https://arxiv.org/abs/1406.2661v1}
\BIBentrySTDinterwordspacing

\bibitem{Goodfellow2016NIPSNetworks}
\BIBentryALTinterwordspacing
I.~Goodfellow, ``{NIPS 2016 Tutorial: Generative Adversarial Networks},''
  \emph{Neural Information Processing Systems}, 2016. [Online]. Available:
  \url{http://www.iangoodfellow.com/slides/2016-12-04-NIPS.pdf}
\BIBentrySTDinterwordspacing

\bibitem{Zhu2017UnpairedNetworks}
\BIBentryALTinterwordspacing
J.-Y. Zhu, T.~Park, P.~Isola, and A.~A. Efros, ``{Unpaired Image-to-Image
  Translation using Cycle-Consistent Adversarial Networks},'' \emph{Proceedings
  of the IEEE International Conference on Computer Vision}, vol. 2017-October,
  pp. 2242--2251, 3 2017. [Online]. Available:
  \url{https://arxiv.org/abs/1703.10593v7}
\BIBentrySTDinterwordspacing

\bibitem{Li2018Query-EfficientLearning}
P.~Li, J.~Yi, and L.~Zhang, ``{Query-Efficient Black-Box Attack by Active
  Learning},'' 2018.

\bibitem{Zhou2020DaST:Attacks}
M.~Zhou, J.~Wu, Y.~Liu, S.~Liu, and C.~Zhu, ``{DaST: Data-free Substitute
  Training for Adversarial Attacks},'' 2020.

\bibitem{Goodfellow2015ExplainingExamples}
I.~J. Goodfellow, J.~Shlens, and C.~Szegedy, ``{Explaining and Harnessing
  Adversarial Examples},'' 2015.

\bibitem{Rozsa2016AdversarialGeneration}
A.~Rozsa, E.~M. Rudd, and T.~E. Boult, ``{Adversarial Diversity and Hard
  Positive Generation},'' 2016.

\bibitem{Papernot2017PracticalLearning}
N.~Papernot, P.~McDaniel, I.~Goodfellow, S.~Jha, Z.~B. Celik, and A.~Swami,
  ``{Practical Black-Box Attacks against Machine Learning},'' 2017.

\bibitem{Sharafaldin2019DevelopingTaxonomy}
I.~Sharafaldin, A.~H. Lashkari, S.~Hakak, and A.~A. Ghorbani, ``{Developing
  Realistic Distributed Denial of Service (DDoS) Attack Dataset and
  Taxonomy},'' in \emph{2019 International Carnahan Conference on Security
  Technology (ICCST)}, 2019, pp. 1--8.

\bibitem{Kingma2015Adam:Optimization}
\BIBentryALTinterwordspacing
D.~P. Kingma and J.~Ba, ``{Adam: A Method for Stochastic Optimization},'' in
  \emph{3rd International Conference on Learning Representations, ICLR 2015,
  San Diego, CA, USA, May 7-9, 2015, Conference Track Proceedings}, Y.~Bengio
  and Y.~LeCun, Eds., 2015. [Online]. Available:
  \url{http://arxiv.org/abs/1412.6980}
\BIBentrySTDinterwordspacing

\bibitem{LeCun2010MNISTDatabase}
Y.~LeCun, C.~Cortes, and C.~J. Burges, ``{MNIST handwritten digit database},''
  \emph{ATT Labs [Online]. Available: http://yann.lecun.com/exdb/mnist},
  vol.~2, 2010.

\bibitem{Liu2015DeepWild}
Z.~Liu, P.~Luo, X.~Wang, and X.~Tang, ``{Deep Learning Face Attributes in the
  Wild},'' in \emph{Proceedings of International Conference on Computer Vision
  (ICCV)}, 12 2015.

\end{thebibliography}

\end{document}